\documentclass[a4paper,twocolumn,11pt,accepted=2026-10-20]{quantumarticle}
\pdfoutput=1
\usepackage{array}[=2016-10-06]
\usepackage[T1]{fontenc}
\usepackage[numbers,sort&compress]{natbib}
\usepackage{graphicx} 
\usepackage{physics}
\usepackage{dcolumn}
\usepackage{microtype}
\usepackage{tabularx} 
\usepackage{enumitem} 
\usepackage{mathtools}
\usepackage{amsmath}
\usepackage{amsfonts}
\usepackage{float}
\usepackage{bm}
\usepackage[colorlinks=true, hyperindex, breaklinks, linkcolor=blue, urlcolor=blue, citecolor=blue]{hyperref}
\usepackage{subfiles}

\newcommand{\todm}[1]{\ket{#1}\bra{#1}}
\newcommand{\todmstar}[1]{\ket*{#1}\bra*{#1}}
\newcommand{\CROT}{\text{CROT}^{1/{m_i}}}
\newcommand{\vardbtilde}[1]{\tilde{\raisebox{0pt}[0.85\height]{$\tilde{#1}$}}}

\DeclareMathOperator*{\argmax}{arg\,max}

\newcounter{protocol}
\newenvironment{protocol}[1]
{
    \par\addvspace{\topsep}
    \noindent
    \tabularx{\linewidth}{@{} X @{}}
    \hline
    \refstepcounter{protocol}
    \textbf{Protocol \theprotocol} #1 \\
    \hline
}
{ \\
    \hline
    \endtabularx
    \par\addvspace{\topsep}
}

\begin{document}

\title{Hardware-Efficient Entanglement Distillation Using Bosonic Systems }

\date{October 30, 2026}


\author{Shoham Jacoby}
\email{shohamjac@gmail.com}
\affiliation{
 Department of Condensed Matter Physics, Weizmann Institute of Science, Rehovot, Israel
}
\affiliation{ Racah Institute of Physics, The Hebrew University of Jerusalem, Jerusalem 91904, Givat Ram, Israel}
\author{Rotem Arnon-Friedman}
\affiliation{
 Department of Complex Systems, Weizmann Institute of Science, Rehovot, Israel
}
\author{Serge Rosenblum}
\affiliation{
 Department of Condensed Matter Physics, Weizmann Institute of Science, Rehovot, Israel
}

\begin{abstract}
    High-fidelity entanglement shared between distant quantum systems is an essential resource for quantum communication and computation. Entanglement distillation addresses this need by converting multiple noisy Bell pairs into fewer higher-fidelity pairs, using only local quantum operations and classical communication. However, this approach typically requires a substantial overhead in the number of qubits. 
    To bypass this hurdle, we propose to leverage the high-dimensional Hilbert space of a \emph{single} pair of bosonic systems to store a large amount of entanglement, replacing the need for multi-qubit systems.
    To distill entanglement in such a setup, we devise a new entanglement distillation protocol, tailored for bosonic systems. The protocol converts a highly-entangled noisy state between two bosonic systems into a lower-dimensional but high-fidelity entangled state, using only local bosonic operations. We show that our protocol significantly enhances the fidelity of the entangled state in the presence of naturally occurring loss and dephasing errors. 
    Compared to methods relying on multiple Bell pairs, our scheme offers a more hardware-efficient strategy, providing a practical route toward the realization of entanglement distillation.
\end{abstract}

\maketitle

\section{Introduction}
    
   Entanglement between spatially separated quantum systems plays a crucial role in quantum communication \cite{KimbleQuantumInternet2008, Bennet1993Teleportation, QuantumRepeaterJiang}, quantum cryptography~\cite{QuantumCrypto1991,Ekert2014crypto}, and distributed quantum computing \cite{Distributedquantumcomputation1999, Nickerson_2013_topologicalComputingNetwork}. 
    However, establishing and maintaining high-fidelity entanglement between distant parties remains a significant challenge due to inherent noise in the preparation and distribution process that degrades the entangled states.

    Entanglement distillation, also known as entanglement purification, offers a promising solution \cite{Bennett_first,Bennett_1996,Deutsch_1996distillation,pureRepeaters1999, Matsumoto_2003stabilizer_entanglement_distillation}.
    
    Typically, distillation protocols process multiple noisy entangled qubit pairs to extract a smaller number of higher-fidelity pairs \cite{Bennett_first,Bennett_1996,Deutsch_1996distillation} using only local operations and classical communication (LOCC) --- the operations available to the two distant parties, each acting on their own local systems.
    Although entanglement distillation has been demonstrated experimentally \cite{Kalb_2017hanson, reichle2006distillationexperimental}, these experiments have been limited to distilling only one high-fidelity pair from two noisy ones. 
    Achieving a Bell-pair fidelity high enough for useful applications
    requires a large number of initial qubits and operations \cite{Zwerger_2018_finite_resource_ED, Thomas2024_finite_resource}, which poses significant technological challenges and introduces potential points of failure that can further degrade the fidelity.

    \begin{figure}[t]
        \begin{minipage}
        {\dimexpr \linewidth\relax} 
\includegraphics[width= 0.85\linewidth]{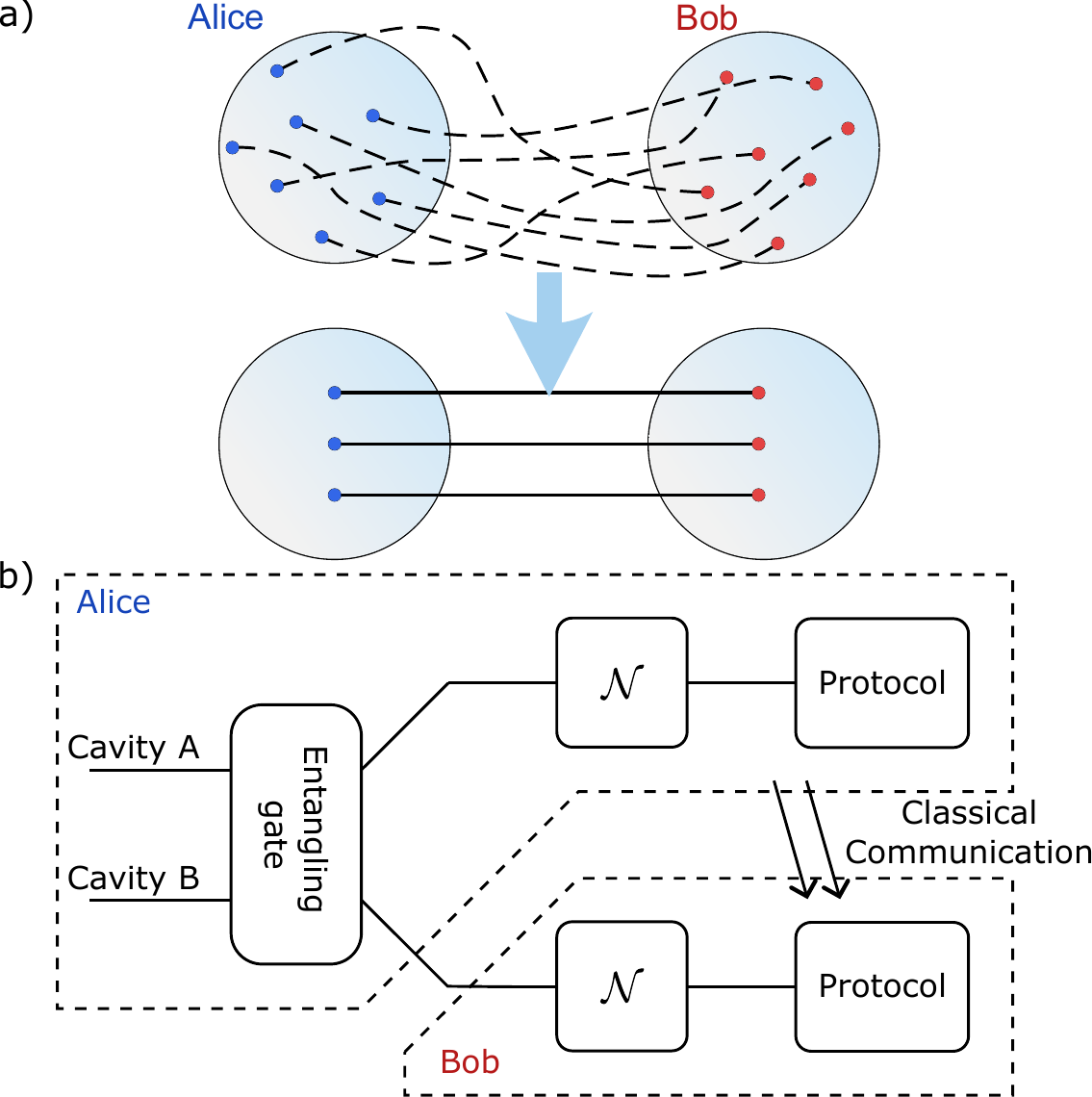} \vspace{-5pt}
    \centering
          \caption{\textbf{(a)}      Top: Initial state of two cavities prepared with high entanglement but low fidelity (represented by dashed lines). 
            Bottom: After applying the entanglement distillation protocol, the state fidelity is enhanced while reducing the total entanglement (solid lines). Distillation is thus achieved
using only a single pair of bosonic systems, eliminating the need for multiple entangled two-qubit systems.
\textbf{(b)} The distillation protocol begins with Alice applying an entangling gate and transferring the quantum state of one cavity mode to Bob's stationary cavity through a lossy channel. During transmission, the cavities experience a loss-dephasing channel $\mathcal{N}$. Subsequently, Alice and Bob use local operations and classical communication to recover a high-fidelity entangled state.
            }
        \label{fig: distillation concept}
        \end{minipage}
    \end{figure}

    To address similar hardware overhead challenges in quantum error correction, bosonic systems have been used as an alternative to two-level systems \cite{Gottesman_2001}. The high-dimensional Hilbert space of bosonic systems enables redundant encoding of logical qubits within a single element, leading to more hardware-efficient error correction protocols  \cite{Gottesman_2001,Cochrane_1999cat_theory,binomial.6.031006,Albert_2018,Grimsmo_2020}. 
    Bosonic quantum error correction has been realized experimentally for a variety of encodings, such as GKP  \cite{Gottesman_2001,fluhmann2019GKPexprimental} and Schr{\"o}dinger cat codes \cite{Mirrahimi_2014, Hu_2019binomexperiment}. The hardware-efficiency of this method has enabled bosonic systems to exceed the break-even point of quantum error correction~\cite{ofek2016parity,Sivak_2023GKP_exp,Campagne_Ibarcq_2020GKPexperiment, ni2023beating, GKP_Quantique}.

 Previous studies on bosonic entanglement distillation have primarily focused on optical systems, where Gaussian states and operations are naturally accessible \cite{takahashi2010entanglement,Rozp_dek_2021,ZhangHybridEntangDist}. However, non-Gaussian operations or resources, which are essential for entanglement distillation \cite{Eisert_2002,Namiki_2014},
remain challenging to implement \cite{Ourjoumtsev_2007}. In contrast, recent hardware developments in superconducting quantum circuits~\cite{ofek2016parity,gertler2021protecting, Milul, Konno_2024, rojkov2024stabilization, ding2024quantumcontroloscillatorkerrcat} have made it increasingly practical to realize certain non-Gaussian operations with high fidelity. 
Additionally, significant advances have been made in distributing entanglement across microwave links, a key requirement for networking quantum processors \cite{Wallraff_link,axline2018demand_Schoelkopf,LinkBurkhart.2.030321,almanakly2024deterministicremoteentanglementusing,storz2023loophole}.
Together, these developments open the door to entanglement distillation protocols explicitly tailored to superconducting quantum circuits.

In this work, we propose to use bosonic systems for achieving hardware-efficient entanglement distillation.
    The high degree of entanglement between the bosonic systems is achieved by populating multiple energy levels of the high-dimensional bosonic Hilbert space. 
    We devise a novel protocol that performs entanglement distillation using only a \emph{single} pair of bosonic systems, rather than relying on multiple copies of entangled two-qubit systems. 
    Similar to qubit-based schemes, the initial state used in our protocol is in principle straightforward and natural to produce in experimental settings. Our distillation protocol then acts on the noisy initial state and produces a final state with enhanced fidelity while reducing the total effective size of the Hilbert space. This setup is illustrated in Fig. \ref{fig: distillation concept}a.

   Our proposed distillation protocol is markedly different from traditional multi-qubit entanglement distillation. While it is true that a bosonic system can be thought of as a tensor product of multiple two-level systems, the native errors and gates on the bosonic system do not respect this tensor-product structure. Thus, standard entanglement distillation protocols used for qubits are not appropriate. Instead, a more adequate approach is to treat the cavities as qudits encoded in a basis that is natural to bosonic systems.
    In particular, we use discrete rotation symmetry in cavity phase space to define the encoding, inspired by the bosonic rotation codes introduced by Grimsmo et al.~\cite{Grimsmo_2020}. This choice is motivated by the effectiveness of these quantum error-correcting codes against naturally occurring loss and dephasing noise.

    We show both analytically and numerically that our method enables entanglement distillation with high performance while minimizing the amount of necessary hardware.
    
    The rest of the paper is organized as follows. In Section~\ref{sec:bed}, we introduce an approximate model for noisy bosonic systems and present our entanglement distillation protocol within this simplified setting. We analyze its performance both analytically and numerically.
    Then, in Section~\ref{sec: bosonic-main}, we transition to a more realistic scenario and propose a practical procedure for implementing the protocol in a two-cavity system using squeezed states manipulated by controlled rotation gates \cite{Grimsmo_2020}, which are operations that are naturally suited to circuit-QED platforms. We also identify the break-even point at which the protocol outperforms the trivial single-photon encoding.

\begin{figure}[t] 
    \centering
    \begin{minipage}{\dimexpr \linewidth\relax} 
        \centering
        \includegraphics[width=\linewidth]{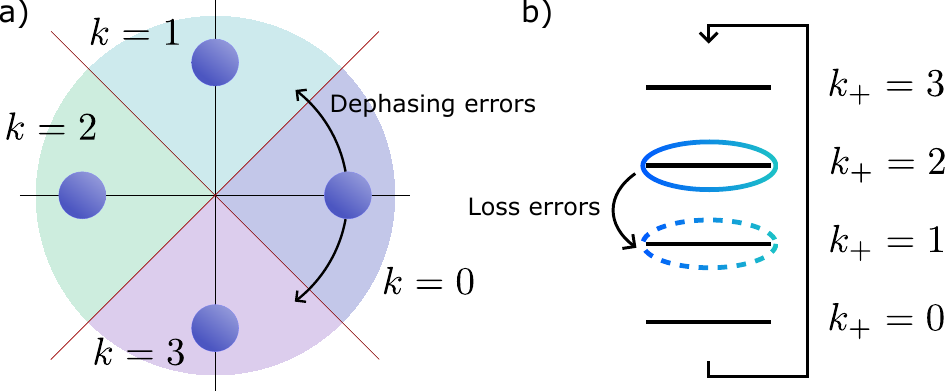}
        \end{minipage}
        \caption{\textbf{(a)} Each qudit is encoded in a bosonic system by dividing its phase space into $d$ angular sections. In this computational basis, dephasing errors correspond to the application of cyclic raising or lowering operators. \textbf{(b)} The dual qubit basis corresponds to a modular photon-number representation, with photon loss acting as a cyclic lowering operator.}
        \label{fig: slicing the cavity}
\end{figure}

 \section{Bosonic Entanglement distillation}\label{sec:bed}

    As illustrated in Fig. \ref{fig: distillation concept}b, the distillation process begins with one party, Alice, preparing a state with high entanglement entropy $E_i$ \cite{Bennett_1996} in two bosonic systems, henceforth referred to as cavities $A$ and $B$. 

        Following this state preparation, Alice transmits the quantum state from cavity $B$
    to Bob. During transmission, both cavities are subject to photon loss and dephasing, which are common noise channels in bosonic systems \cite{Leviant_2022}. Photon gain errors occur at a significantly lower rate in typical experimental settings and are therefore not considered in this study. Nevertheless, our protocol can be straightforwardly extended to incorporate these errors as well.
    The loss-dephasing noise channel reduces the fidelity of the two-cavity state to the ideal entangled state.
    
    To restore a high-fidelity state, Alice and Bob apply an entanglement distillation protocol.
    The general form of the protocol is as follows: first, each party conducts specific measurements on their respective cavities. Alice then communicates her measurement outcomes to Bob. Based on these outcomes, both parties apply bosonic operations on their cavities, with an optional decision rule to determine whether to abort the protocol or not.
    The end result of this protocol is a distilled state that has higher fidelity to a desired target state, albeit with reduced entanglement entropy $E_f$.
    In the next sections, we elaborate on each stage of the distillation procedure.

\subsection{Encoding highly entangled bosonic states}
      
    We define the bosonic qudit by dividing the cavity phase space into $d$ angular sections, and associating the qudit state $\ket{k}$ with the $k^\textrm{th}$ angular interval $\frac{\qty(2k-1)\pi}{d} \le \phi< \frac{\qty(2k+1)\pi}{d}$ (Figure \ref{fig: slicing the cavity}a). A definition using continuous variables is provided in Section \ref{section: Creating the initial state}. For simplicity, we choose $d$ to be a power of two, i.e., $\log d \in \mathbb{N}$.

    We can now introduce our highly entangled initial state
    \begin{align} \label{eq: initial state}
        \ket{\psi_i} = \frac{1}{\sqrt{m_i}}\sum_{k=0}^{m_i-1} \ket*{\frac{d}{m_i} k}_A \otimes \ket*{\frac{d}{m_i} k}_B \;,
    \end{align}

    where $m_i$ is the initial number of ``legs'', i.e. the number of basis states participating in the superposition. As we show in Section \ref{section: Creating the initial state}, although the ideal initial state requires infinite energy to confine qudit states to a single angular section, it can be well approximated by finite-energy bosonic states that are convenient to prepare.
    The entanglement entropy between the cavities at the onset of the distillation protocol is thus $E_i \equiv H\qty(\Tr_A\todm{\psi_i})=\log{m_i}$, where $H\qty(\rho)\equiv-\Tr(\rho \log \rho)$ is the von Neumann entropy of a state $\rho$, and $\Tr_{A}$ denotes the partial trace over subsystem $A$. For simplicity, we assume $m_i=d$ unless stated otherwise.

 \subsection{Discretized error model} \label{sec: discrete error model}

\begin{figure*}[t] 
    \begin{protocol}{Bosonic entanglement distillation \label{pro: general protocol}} 
        \textbf{Arguments:}
            \begin{itemize}[itemsep=-1mm]
                \item $m_i$ -- The dimension of the single-cavity subspace supporting the initial state, corresponding to the angular phase-space symmetry of the initial state.
                \item $m_c$ -- The dimension of the single-cavity subspace supporting the intermediate qudit obtained after the first measurement stage.
                \item $m_f$ -- The dimension of the single-cavity subspace supporting the final qudit obtained after both measurement stages.
                \item $f_\text{cut}\in\qty{0}\cup[1/{m_f^2},1)$ -- The cut-off fidelity below which the protocol is aborted.
                \item all dimensional arguments are restricted to powers of two ($\log{m_x}\in \mathbb{N}$) and follow the relation: $2|m_f|m_c|m_i|d$, where $|$ stands for ``divides".
            \end{itemize}
        \begin{enumerate}[label=\arabic*:, itemsep=-1mm, leftmargin=*, ref=\arabic*]
            \item \textbf{Both parties} perform modular $\Delta_c \equiv \frac{d}{m_c}$ measurements in their qudit's computational basis and register their results as $A_1$ and $B_1$, respectively.
            \item \textbf{Alice} sends her result to Bob and \textbf{Bob} determines the most likely rotation angle $u_B$ using a decoder based on both results (Eq. \eqref{eq:counting_rotation}).
            \item \textbf{Alice and Bob} rotate their states by $X^{-A_1}$ and $X^{-B_1+u_B\Delta_c}$, respectively.
            \item \textbf{Both parties} perform modular $\Delta_f \equiv \frac{m_c}{m_f}$ measurements in their qudits' dual basis and register their results as $A_2$ and $B_2$, respectively.
            \item \textbf{Alice} sends her result to \textbf{Bob}, who uses a decoder to determine the most likely rotation angle $v_B$ (Eq. \eqref{eq:counting_loss}).
            \item \textbf{Alice and Bob} rotate their states in the dual basis, applying $Z^{-A_2}$ and $Z^{-B_2 - v_B \Delta_f}$, respectively.
            \item \textbf{Abort} if the calculated expected fidelity $\mathcal{F}\qty(A_1,B_1,A_2,B_2) \le f_\text{cut}$ (Eq. \eqref{eq: fidelity specific}).
        \end{enumerate}
    \end{protocol}
    \end{figure*}

   While transferring cavity $B$ from Alice to Bob, both cavities are affected by dephasing and photon loss channels (see Appendix \ref{sec: The loss-dephasing channel}). Because these channels commute~\cite{Leviant_2022}, we can assume they act consecutively on the input state. The dephasing channel results in a rotation of the cavity phase space, leading to cyclic shift errors in the computational (angular) basis (see Figure \ref{fig: slicing the cavity}a). The resulting discretized dephasing errors $D_s$, which shift the state by $s$ angular sections, are equivalent to the generalized Pauli operators \cite{Gottesman_2001} $X^s$, with $-d/2<s \le d/2$, i.e.:
    \begin{align}
         D_s \equiv X^s : & \ket{k} \rightarrow \ket{k+s ~ \qty(\text{mod}~d)} \;.
    \end{align}

     The corresponding dephasing error probabilities $\qty{p_{s}^D}$, which decay super-exponentially with $\abs{s}$, can be derived using the rate of the bosonic dephasing channel,~$\gamma_\phi$ (see Appendix \ref{sec: discrete errors}).
        
    The loss channel is more conveniently expressed in the dual modular photon-number basis (Figure \ref{fig: slicing the cavity}b), whose basis states $\left|k_{+}\right\rangle $ are Fourier transforms of the $d$ computational states
    \begin{align}
            \left|k_{+}\right\rangle \equiv \frac{1}{\sqrt{d}} \sum_{j=0}^{d-1}e^{-2ijk\pi/d}\left|j\right\rangle  \;.
    \end{align}
    In this basis,
    the loss of $l$ photons can be approximately modeled by a cyclic shift, which is equivalent to the generalized Pauli operator $Z^{-l}$, where $l \ge 0$:
    \begin{align}
        L_l \equiv Z^{-l} : & \ket{k_+} \rightarrow \ket{\qty(k-l)_+ ~ \qty(\text{mod}~d)} \;.
    \end{align}
    The corresponding error probabilities $\qty{p^L_{l}}$ can be approximated using the average photon number $\bar{n}$ and the rate $\gamma_l$ associated with the bosonic loss channel (Appendix \ref{sec: discrete errors}). These probabilities also decrease super-exponentially with $l$.
    Henceforth, the modulo symbol will be omitted for brevity.

This discretized qudit error model, while not exact, is a reasonable approximation for our purposes, as we show numerically in Section \ref{subsection: Continuous vs. discrete and the effect of squeezing}.
For two cavities undergoing both loss and dephasing noise, we define the set of possible errors, $\mathcal{E}$, and calculate the corresponding probability $p\qty(\bar{x}) = p^D_{s_A} p^D_{s_B} p^L_{l_A} p^L_{l_B}$ of each error $\bar{x} = \qty(s_A,s_B,l_A,l_B) \in \mathcal{E}$.
Applying this noise channel to the initial state from Equation \eqref{eq: initial state}, and considering all possible errors, we obtain the density matrix

\begin{align} \label{eq: initial noisy state}
    \rho &= \sum_{\mathclap{\bar{x} \in \mathcal{E}}}
    p\qty(\bar{x})\todm{\varphi_{\bar{x}}} \;,
\end{align}
where 
$\ket{\varphi_{\bar{x}}}$
is the error state associated with $\bar{x}$: 
    \begin{multline} \label{eq: pure noisy state}
        \ket{\varphi_{\bar{x}}} 
        =
        L^A_{l_A} L^B_{l_B}
        D^A_{s_A} D^B_{s_B}  \ket{\psi_i} 
        \\
        =
        L^{A}_{l_A} L^{B}_{l_B} \sum_{k=0}^{m_i-1} \ket{k}_A \ket{ k + \delta_s}_B \;,
    \end{multline}
    with $\delta_s = s_B -s_A$.

    \begin{figure*}[t] 
        \begin{minipage}{\dimexpr\linewidth\relax} 
            \includegraphics[width=\linewidth]{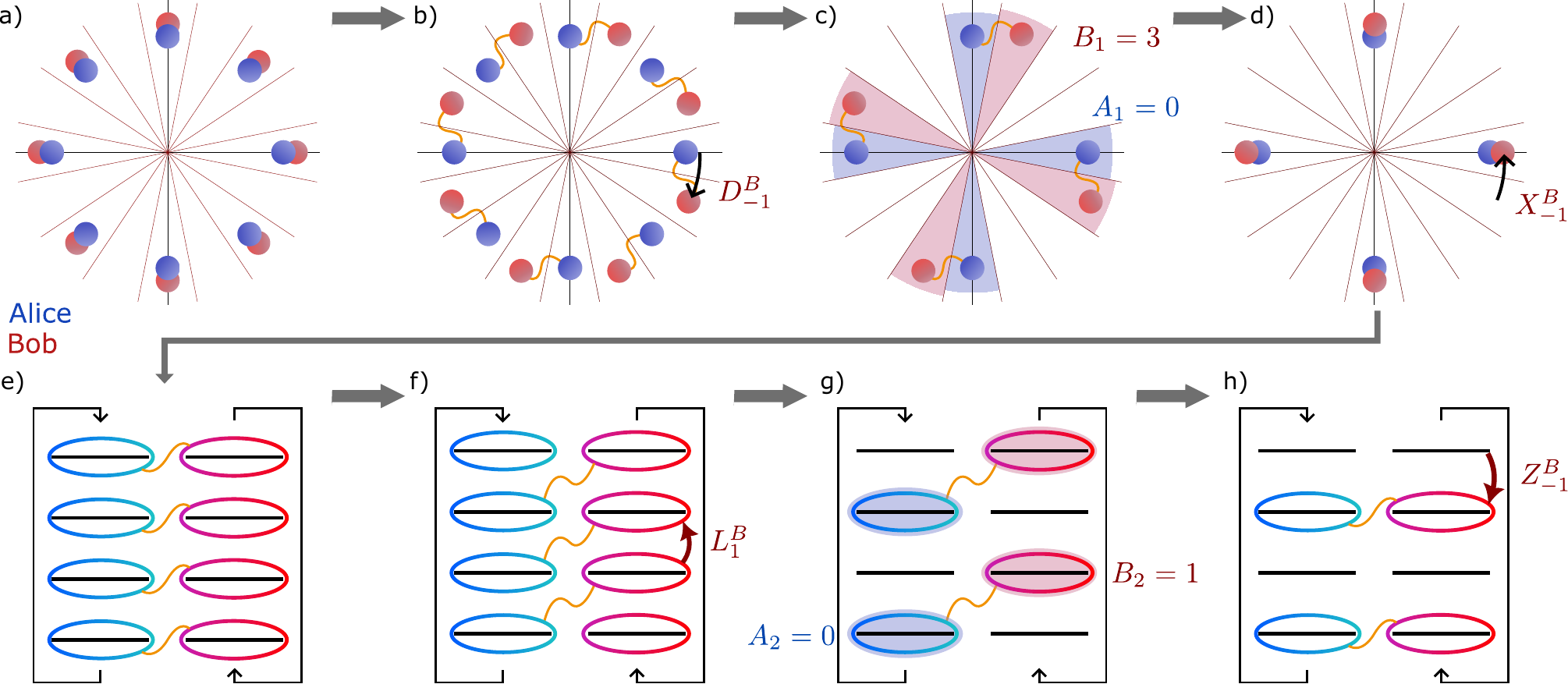}
        \end{minipage}
        \caption{Schematic representation of the entanglement distillation protocol with $d=16$, $m_i=8$, $m_c=4$, and $m_f=2$. Alice's and Bob's phase space distributions are illustrated in blue and red, respectively. 
        Here, we choose $m_i \neq d$ for improved clarity.
        \textbf{(a)} The protocol starts with the two cavities in the initial entangled state. \textbf{(b)} The cavities then undergo dephasing noise. In this example, only Bob's cavity is affected by $D^B_{-1}$. \textbf{(c)} Next, both parties carry out a modular phase measurement with $m_c=4$, which projects the state onto a four-dimensional subspace for each cavity. \textbf{(d)} Next, the parties share their measurement outcomes. In this example, only Bob rotates his cavity phase space to realign with that of Alice.
        The second stage of the protocol, which addresses photon loss, is illustrated in \textbf{(e-h)}. Alice's and Bob's cavities are depicted in the modular photon-number basis on the left and right sides, respectively. Due to the negative sign in Bob's cavity (see Eq. \eqref{eq: State for second step}), photon loss results in shifts in opposite directions, as indicated by the arrows representing photon-number modularity. Similarly to the top panel, the bottom panel represents \textbf{(e)}, the initial noiseless cavities, \textbf{(f)}, the cavities after undergoing the photon loss channel, \textbf{(g)}, the modular photon-number measurements, and \textbf{(h)}, realignment by the parties after communicating the measurement outcomes.
        }
        \label{fig: the protocol}
    \end{figure*}
\subsection{Entanglement distillation protocol} \label{sec: Entanglement distillation protocol}

After both cavities undergo the loss-dephasing channel, Alice and Bob apply the distillation protocol using LOCC operations to achieve a high-fidelity entangled state (see Protocol \ref{pro: general protocol}). In the following sections, we present the protocol in a general setting, followed by an example with a four-level qudit encoding.

In the first of three stages of the protocol (Fig. \ref{fig: the protocol}c-d), Alice and Bob measure their cavities with modular phase measurements to address dephasing errors. In the second stage (Fig. \ref{fig: the protocol}g-h), they perform modular measurements in the dual number basis to address photon loss errors. Finally, in a third
stage, the parties decide whether or not to abort the protocol, based on
the measurement outcomes.

\textit{Modular phase measurements} -- The modular phase measurement is represented by the set of projection-valued measurement elements (PVMs), $\qty{P_x^{\Delta_c}}_{x=0}^{x= \Delta_c-1}$, where $x$ is the measurement outcome corresponding to the angular section modulo $\Delta_c$, with $\Delta_c$ a power of two between $1$ and $m_i$. 
Each PVM element can be written as
\begin{align} \label{eq: modular phase measurement}
P_x^{\Delta_c} = \sum_{j=0}^{\mathclap{d/\Delta_c-1}} \todm{ j\Delta_c  + x}.
\end{align}

This measurement restricts the qudit from the initial subspace of dimension $m_i$ to a subspace of dimension $m_c\equiv d/\Delta_c$ (see Fig. \ref{fig: the protocol}c). $m_c$ is a key parameter of the protocol. A lower $m_c$ will result in higher robustness against dephasing errors, while a higher $m_c$ will result in higher robustness against loss errors; a detailed discussion of this trade-off will be presented in Section \ref{subsection: Performance of the protocol}. We note that increasing $m_i$ while keeping $m_c$ fixed does not enhance protection against dephasing errors (Appendix \ref{sec: entanglement resource}). However, as we show in Section \ref{section: Creating the initial state}, choosing a higher $m_i$ can lead to improved state preparation fidelity.
     
    Alice and Bob each apply a modular phase measurements to their respective cavity, yielding outcomes $A_1$ and $B_1$. These measurements project the $m_i$-legged state to an $m_c$-legged state (with $m_c\leq m_i$) rotated by $A_1\left(B_1\right)$ sections relative to the state $\ket{0}_{A(B)}$. 
    Due to the dephasing noise, the measurement outcomes $A_1$ and $B_1$ will differ by $\delta_s\,\text{mod}\,\Delta_c$.
    Using Eqs.\,\eqref{eq: pure noisy state} and \eqref{eq: modular phase measurement}, we obtain the post-measurement state
    \begin{multline} \label{eq: collapsed state}
        \ket{\tilde{\varphi}_{\bar{x}}}
        \\
        = \frac{1}{\sqrt{m_c}} \sum_{k'=0}^{m_c-1} c_{k'} \ket{\Delta_c k' + A_1}_A \ket{\Delta_c \qty(k' + u) + B_1}_B ,
    \end{multline}
    with $u = \frac{\delta_s + A_1-B_1}{\Delta_c}$ an integer. $\abs{c_{k'}}=1$ accounts for loss errors and can be ignored for the examination of the first stage of the protocol (see Appendix \ref{sec: analyzing the protocol}). 

    This expression highlights two types of misalignment between the cavities: one where each cavity is rotated with respect to the state $\ket{0}_{A(B)}$ by $A_1$($B_1$), and another where the cavities are rotated by $u\Delta_c$ with respect to each other.
    Both parties need to rotate their cavities to address those misalignments, with the objective of achieving the intermediate target state $\frac{1}{\sqrt{m_c}}\sum_{k'} c_{k'} \ket{\Delta_c k'}_A \ket{\Delta_c k'}_B$.
        
    To align her cavity with $\ket{0}_A$, Alice rotates her cavity by applying $X^{-A_1}$. Next, Alice communicates her measurement result to Bob.
    Since the value of the relative rotation, $u \Delta_c$, cannot be determined solely from $A_1$ and $B_1$, Bob uses a maximum-likelihood decoder (see Appendix \ref{sec: analyzing the protocol}) to determine the most likely value $u_B$. He then corrects the cavity state by applying $X^{-B_1+u_B\Delta_c}$, thereby optimizing the fidelity to the intermediate target state. 

    This decoder predicts the most likely value of $u$ by summing the probabilities of all errors $\bar{x}$ that are consistent with the measurement outcomes $A_1$ and $B_1$ for each possible value of $u$.

    We note that the maximum-likelihood decoder need not be evaluated in real time. For fixed protocol parameters and calibrated noise rates, the optimal corrections can be precomputed for every possible pair of measurement outcomes and stored in a lookup table. The online decoding step is therefore only a table lookup. Since the number of angular sections $d$ is limited in practice by the finite phase-space localization of the primitive state, the number of possible outcomes is modest in the parameter regimes considered here.
    
    \textit{Modular number measurements} -- If the rotation was successful in correcting both types of angular misalignment, the state after completing the first stage can be rewritten in the dual basis as
    \begin{align} \label{eq: State for second step}
        & L^A_{l_A} L^B_{l_B} \frac{1}{\sqrt{m_c}} \sum_{k=0}^{m_c-1} \ket{k_+^{m_c}}_A \ket{-k_+^{m_c}}_B
        \\ \label{eq: state for second step expanded}
        & = \frac{1}{\sqrt{m_c}}  \sum_{k=0}^{m_c-1} \ket{\qty(k-l_A)_+^{m_c}}_A \ket{\qty(-k-l_B)_+^{m_c}}_B \;,
    \end{align}
    with 
    $\ket{k_+^{m_c}} \equiv \frac{1}{\sqrt{ d / m_c}} \sum_{j=0}^{d / m_c-1} \ket*{\qty(k + j m_c)_+}$ 
    representing a change in the modularity of the dual photon number basis from $d$ to $m_c$. 
    Equation \eqref{eq: State for second step} resembles the structure of the initial state in the computational basis, but with a negative sign in Bob's cavity (since the total photon number must be an integer multiple of $m_c$).
    
    We will use this similarity in the second, loss correction stage, which is dual to the dephasing correction stage. 
    Both parties measure the photon number modulo $\Delta_f$ as described by a set of PVMs $\qty{P_{x+}^{\Delta_f}}_{x=0}^{x=\Delta_f-1}$, where $x$ is the measurement result. Each PVM element can be written as
    \begin{align} \label{eq: dual-basis modular measurement}
    P_{x+}^{\Delta_f} = \sum_{j=0}^{\mathclap{d/\Delta_f-1}} \todmstar{(j\Delta_f + x)_+} \;.
    \end{align}
    This measurement determines the dimension $m_f\equiv \frac{m_c}{\Delta_f}$ of the subspace supporting the final qudit. 
    Alice and Bob register their measurement outcomes as $A_2$ and $B_2$, respectively. Alice then rotates her cavity in the dual basis by applying $Z^{-A_2}$, and sends her measurement outcome to Bob. 
    Bob rotates his cavity in the dual basis by applying $Z^{-B_2-v_B\Delta_f}$, where $v_B \Delta_f$ is the most likely value for the relative misalignment, which is determined using both outcomes and a maximum-likelihood decoder (Appendix \ref{sec: analyzing the protocol}).
    
    If the rotation in the second stage is successful, we obtain the desired target state
    \begin{align}
        \ket{\psi_t} = \sum_{k=0}^{m_f-1} \ket{\qty(\Delta_f k)^{m_c}_+ }_A \ket{\qty(-\Delta_f k )^{m_c}_+ }_B \;.
    \end{align}
    As in the first stage, a rotation that does not successfully realign the states leads to an erroneous state and a reduced fidelity of the protocol.
    While the entanglement entropy of the target state $E_f = \log{m_f}< E_i$ is reduced by the protocol, we demonstrate in Section \ref{subsection: Performance of the protocol} through numerical analysis that the fidelity of the distilled state to the ideal target state is significantly increased.

    \textit{Conditional abort} -- The fidelity of the distilled state can be further improved by performing a final step in the protocol, which requires two-way communication. In this step, Bob calculates the expected fidelity conditioned on the measurement results $\mathcal{F}\qty(A_1,B_1,A_2,B_2)$ and decides to abort the protocol if the result is lower than a predetermined cut-off $f_\mathrm{cut} \in \qty{0}\cup[1/m_f^2, 1)$, where $1/{m_f^2}$ is the fidelity of the maximally mixed state.
    Setting a higher $f_\text{cut}$ improves the average fidelity of the protocol but also increases the abort probability $p_\text{abort}$. 
    Choosing \( f_\text{cut}=0 \) corresponds to omitting this step, making two-way communication unnecessary. However, this choice also reduces the protocol's effectiveness (see Section \ref{sec: Aborting and restarting the protocol}).
    \vspace{7pt}

    We note that our protocol can be interpreted as a stabilizer-based entanglement distillation scheme \cite{Matsumoto_2003stabilizer_entanglement_distillation}, with a stabilizer group generated by $\langle  Z^{m_c}, X^{d/m_f} \rangle$.

    \vspace{7pt}
    
    To benchmark the performance of the distillation protocol, we compare it to a scenario in which the parties do not communicate. 
    The optimal strategy without relying on classical communication involves each party merging adjacent angular sections in phase space to reduce dephasing errors and merging adjacent dual-basis states to minimize loss errors (see Appendix \ref{sec: no-communication}). This strategy effectively removes errors in redundant degrees of freedom and can, therefore, be interpreted as a form of error correction. 
    We show in Section \ref{subsection: Performance of the protocol} that the performance of our distillation protocol exceeds that of the no-communication protocol.

    \subsection{Four-level example} \label{sec:four-level-example}

    \begin{figure*}[t] 
        \begin{minipage}{\dimexpr\linewidth\relax} 
            \includegraphics[width=\linewidth]{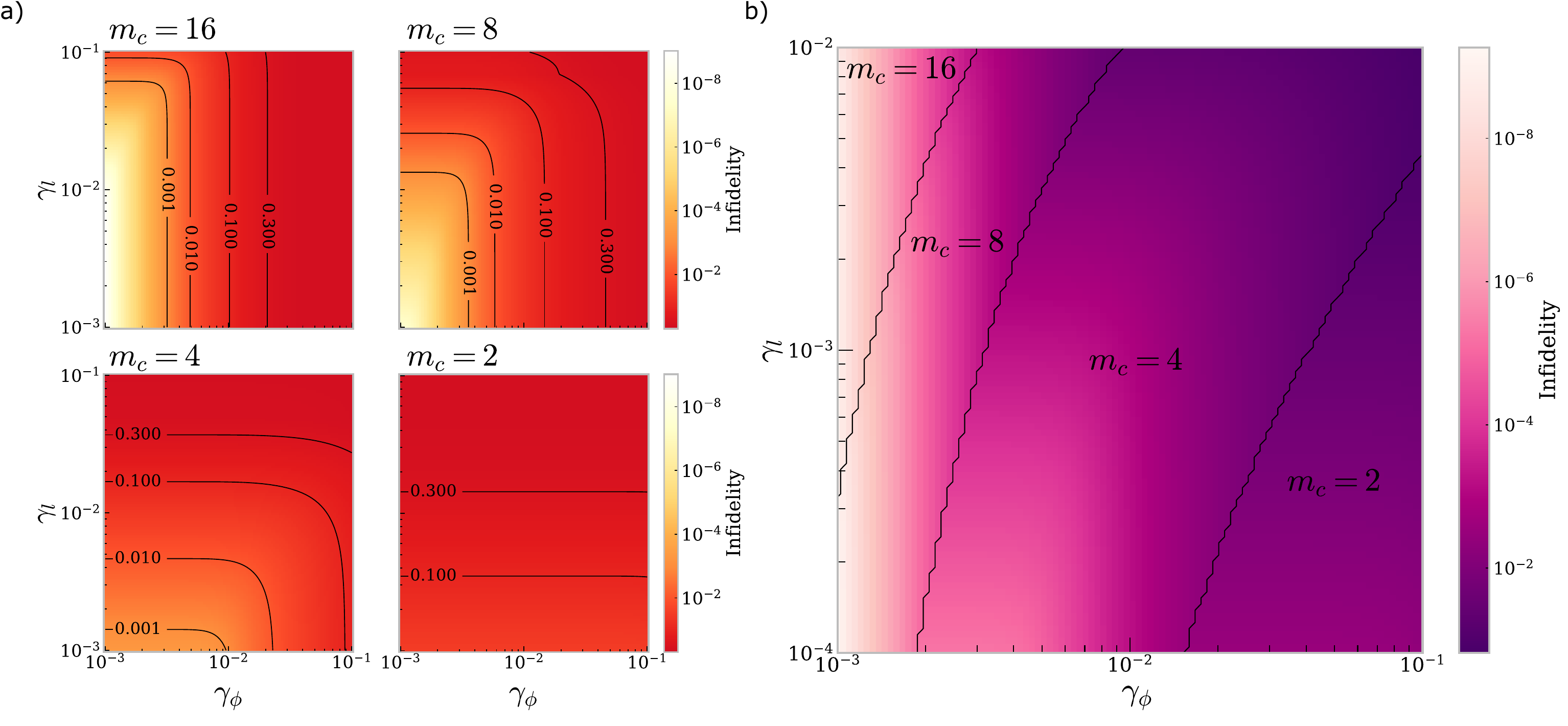}
        \end{minipage}
        \caption{Infidelity of the distilled state using the discrete model with initial qudit dimension $d=m_i=16$ and average photon number $\bar{n}=16$. \textbf{(a)} Infidelity as a function of both error rates $\gamma_l$ and $\gamma_\phi$ for different intermediate qudit dimensions $m_c$. 
        The results indicate that the distilled state fidelity is optimized by selecting different values of $m_c$ for each error rate.
        \textbf{(b)} Infidelity of the distilled state when choosing the optimal $m_c$ for each set of error rates $\gamma_l$ and $\gamma_\phi$. The black lines separate between regions with different optimal $m_c$.
        }
        \label{fig: numerical results}
    \end{figure*}

   To illustrate the distillation procedure, we examine the protocol for a simplified model: a four-level system with $d=m_i=4$.
    We choose $m_c=m_f=2$, indicating that only the dephasing correction stage is applied. Therefore, we do not consider loss errors in this example and set $\gamma_l=0$. The cut-off parameter is chosen to be $f_\text{cut}=0.5$.

    We consider the initial entangled state $\ket{\psi_i} = \frac{1}{\sqrt{4}} \sum_{i=0}^4 \ket{k}_A \otimes \ket{k}_B  = \frac{1}{\sqrt{4}}\qty(\ket{00}+\ket{11}+\ket{22}+\ket{33}) $, with $\ket{kl}\equiv \ket{k}_A\otimes\ket{l}_B$ (see Eq.~\eqref{eq: initial state}) and with initial entanglement entropy $E_i=2$. The target state is $\ket{\psi_t}=\frac{1}{\sqrt{2}}\qty(\ket{00}+\ket{22})$, which has an entanglement entropy of $E_f=1$.
    The state is affected by the dephasing channel $\mathcal{N}_D$ and evolves to the noisy state (see Eq.~\eqref{eq: initial noisy state})
    \begin{multline}
         \rho_{d=4} = p_0\todm{\psi_i}+p_{1}\todm{\varphi_{+}}
         \\
         +p_{1}\todm{\varphi_{-}}+p_{2}\todm{\varphi_{\circ}} \;,
     \end{multline}
     where
     \begin{align*}
        \ket{\varphi_{+}} & = \frac{1}{\sqrt{4}}\left(\ket{0 1} +\left|12\right\rangle +\left|23\right\rangle +\left|30\right\rangle \right)
        \\
        & \quad \quad \quad = D_B^1\ket{\psi_i} = D^{-1}_A\ket{\psi_i} \;,
        \\
        \ket{\varphi_{-}} & =\frac{1}{\sqrt{4}}\left(\left|0 3\right\rangle +\left|10\right\rangle +\left|21\right\rangle +\left|32\right\rangle \right) 
        \\
        & \quad \quad \quad = D_A^1\ket{\psi_i} = D^{-1}_B\ket{\psi_i}  \;,
        \\
        \ket{\varphi_{\circ}} & =\frac{1}{\sqrt{4}}\left(\ket{02} +\ket{13} +\left|20\right\rangle +\left|31\right\rangle \right) 
        \\
        & \quad \quad \quad = D_A^{\pm2}\ket{\psi_i} = D^{\pm2}_B\ket{\psi_i} = D_B^{\pm1} D_A^{\mp1}\ket{\psi_i}.
     \end{align*}
    The states $\ket{\varphi_\pm}$ are associated with a single dephasing error and occur with a probability $p_1$ 
 (see Appendix \ref{sec: discrete error model}). $\ket{\varphi_\circ}$ is the state after two dephasing errors, occurring with a probability $p_2$. The probability of the system remaining in its initial state is denoted by $p_0=1-2p_1-p_2$. We assume small error rates, such that $p_0\gg p_1 \gg p_2$.

    The parties next measure the modular phase, which,  for $m_c=2$, corresponds to a
    parity measurement in the computational basis. Alice and Bob record their outcomes as $A_1$ and $B_1$, respectively. 
    
    If the parties obtain different parities, i.e., $A_1 \neq B_1$, they conclude that the original state was either $\ket{\varphi_{+}}$ or $\ket{\varphi_{-}}$ with equal probability. The party with the odd measurement result rotates its cavity (step 3 in Protocol \ref{pro: general protocol}), aligning it with the $\ket{0}$ state.
    The resulting, unnormalized,  state is
    \begin{align}
        \rho_{\qty(A_1 \neq B_1)} 
        = p_1 \qty(\todm{\widetilde{\varphi_+}} + \todm{\widetilde{\varphi_-}}) \;,
    \end{align}
    with
    $\ket*{\widetilde{\varphi_+}} = 
    \frac{1}{\sqrt{2}}\qty(\ket{00}+\ket{22})
    = \ket{\psi_t}$
    and
    $\ket*{\widetilde{\varphi_-}} = \frac{1}{\sqrt{2}} \qty(\ket{02} + \ket{20})$.
    When obtaining these measurement outcomes, the two parties  conclude that the fidelity of the final state relative to the target state is $0.5$, which is equal to $f_\text{cut}$. Therefore, Alice and Bob abort the protocol. 
    
    In contrast, for $A_1=B_1$, the measured state must have originated from either $\ket{\psi_i}$ or $\ket{\varphi_\circ}$. 
    When both outcomes are even, i.e., $A_1=B_1=0$, the systems are already aligned with $\ket{0}$ and the parties do not need to perform rotations. For $A_1=B_1=1$, the parties rotate their cavities to realign their states with the $\ket{0}$ state. In either scenario, the final state is
    \begin{align}
        \sigma_{f} = p_0\todmstar{\psi_t}+p_2\todm{\widetilde{\varphi_\circ}} \;,
    \end{align}
    where 
    $\ket{\widetilde{\varphi_\circ}} 
    = \frac{1}{\sqrt{2}} \qty(\ket{02} + \ket{20})$. The fidelity of the distilled state to the desired target state $\ket*{\psi_t}$ is therefore $\frac{p_0}{\qty(p_0+p_2)}$.

    Due to the super-exponential nature of the noise channel, we obtain $p_2 \ll p_1^2$. This unique feature of bosonic noise leads to a much more efficient suppression of errors with the bosonic distillation protocol compared to the corresponding qubit distillation protocols.

    We next benchmark the distillation protocol performance by comparing it to the no-communication protocol. Merging adjacent angular sections in phase space effectively coarse-grains the qudits to qubits. In the case of a traditional distillation protocol using two Bell pairs of qubits, this procedure is equivalent to discarding one of the Bell pairs.
    The coarse-graining procedure results in a fidelity to the target state of $1 - p_1 - p_2$, which is lower than our distillation protocol fidelity when $p_2 < \frac{1}{2}\qty(1-2 p_1)$. This inequality holds for the regime of interest, in which $p_0\gg p_1 \gg p_2$.

\subsection{Performance of the protocol} \label{subsection: Performance of the protocol}

    Next, we numerically analyze the performance of the protocol for higher-dimensions, where analytical derivations become impractical. In Figure \ref{fig: numerical results}a, we show the simulated infidelity between the distilled state and the target state as a function of the loss and dephasing rates $\gamma_l$, $\gamma_\phi$. We plot the infidelity for four different values of the intermediate dimension parameter, $m_c$, assuming $m_i=d=16$ and $m_f=2$. We also choose $f_\mathrm{cut}=0$, meaning the protocol is never aborted. 

    We observe that when the dephasing rate is dominant, smaller values of $m_c$ yield a higher fidelity (see \ref{fig: numerical results}b). Conversely, when loss errors dominate, choosing a larger $m_c$ improves the fidelity. This behavior can be explained by noting that a smaller $m_c$ increases the separation between angular sections supporting the state after the first distillation stage, corresponding to a higher tolerance to dephasing errors. When loss is dominant, choosing a higher $m_c$ increases the separation in the dual basis after the second distillation stage, increasing the robustness to loss errors.

    The trade-off between loss and dephasing distillation can also be understood by noting that the first distillation stage maps the state from a space of dimension $m_i^2$ to a subspace of dimension $m_c^2$ to correct dephasing errors, followed by a map to a subspace of dimension $m_f^2$ to correct loss errors. This is akin to distilling the entanglement of $\log m_i$ qubit pairs in traditional multi-qubit distillation to $\log m_c$ qubit pairs and then to $\log m_f$ qubit pairs in a second step. This comparison suggests that $m_c$ controls the balance between the two stages, allowing us to optimize the final fidelity.
    In Figure \ref{fig: numerical results}b, we show the highest achievable fidelity for each pair $\qty(\gamma_l, \gamma_\phi)$ by maximizing over $m_c$. 

\begin{figure}[t] 
    \begin{minipage}{\dimexpr\linewidth\relax} \includegraphics[width=\linewidth]{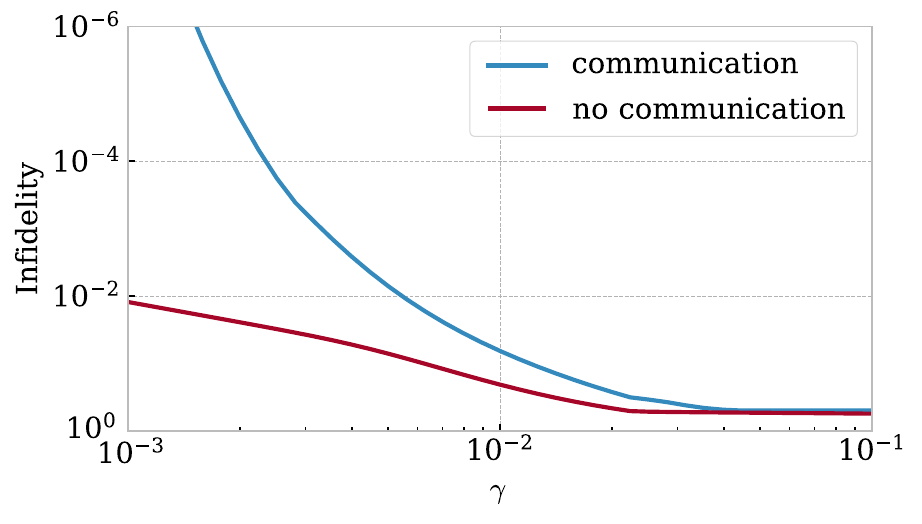}
        \end{minipage}
        \caption{Infidelity of the distilled state when performing the protocol with communication (blue line) and without communication (red line), with $m_i=16,m_f=2,\bar{n}=49$ and $\gamma=\gamma_l=\gamma_\phi$. In the distillation protocol, we maximize the fidelity by choosing the optimal $m_c$, whereas in the no-communication protocol, the optimal rotation strategy is chosen for each error rate.}
        \label{fig: no-communication}
    \end{figure}

    As discussed in Section \ref{sec: Entanglement distillation protocol}, we can evaluate the performance of the distillation protocol by comparing it to a no-communication strategy.
    In this approach, the parties must agree on a predetermined strategy in advance, where each party applies operations to their cavity based solely on their own measurement outcomes.
    Figure \ref{fig: no-communication} compares the infidelity for both strategies, showing that the distillation protocol outperforms the no-communication strategy across the considered parameters. 

   The protocol can also be used to distill states with higher final entanglement entropy. To achieve this, we can set the final cavity state dimension, $m_f$, to be greater than two at the cost of a reduction in the protocol fidelity. The performance of this scenario is analyzed in Appendix \ref{sec: m_f neq two}.

        \begin{figure}[t] 
    \begin{minipage}{\dimexpr\linewidth\relax} 
    \includegraphics[width=\linewidth]{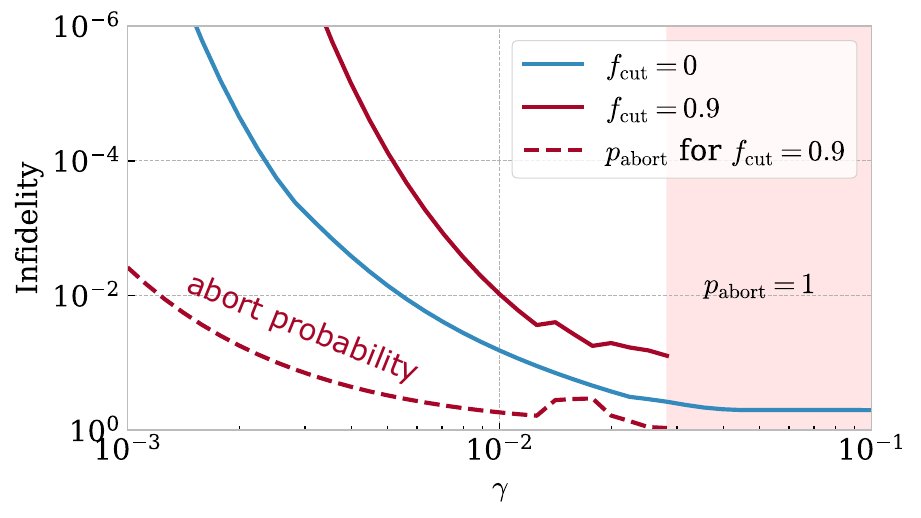}
        \end{minipage}
        \caption{Infidelity of the distilled state when performing the protocol with abort criterion (red line, $f_\text{cut}=0.9$) and without abort criterion (blue line, $f_\text{cut}=0$), for $d=32,m_i=16,m_f=2,\bar{n}=49$ and $\gamma_l=\gamma_\phi = \gamma$. In both scenarios, we set  $m_c$ to optimize the fidelity. 
        This can lead to non-monotonic behavior of the abort probability (dashed line).
        The non-monotonic behavior of the infidelity is due to low-fidelity states passing the cut-off as $\gamma$ decreases, thereby increasing the average infidelity.
        The shaded region indicates error rates where the fidelity is below $f_\text{cut}$ for all instances of the protocol.}
        \label{fig: aborting}
    \end{figure}

            \begin{figure*}[t] 
        \begin{minipage}{\dimexpr\linewidth\relax} 
            \includegraphics[width=\linewidth]{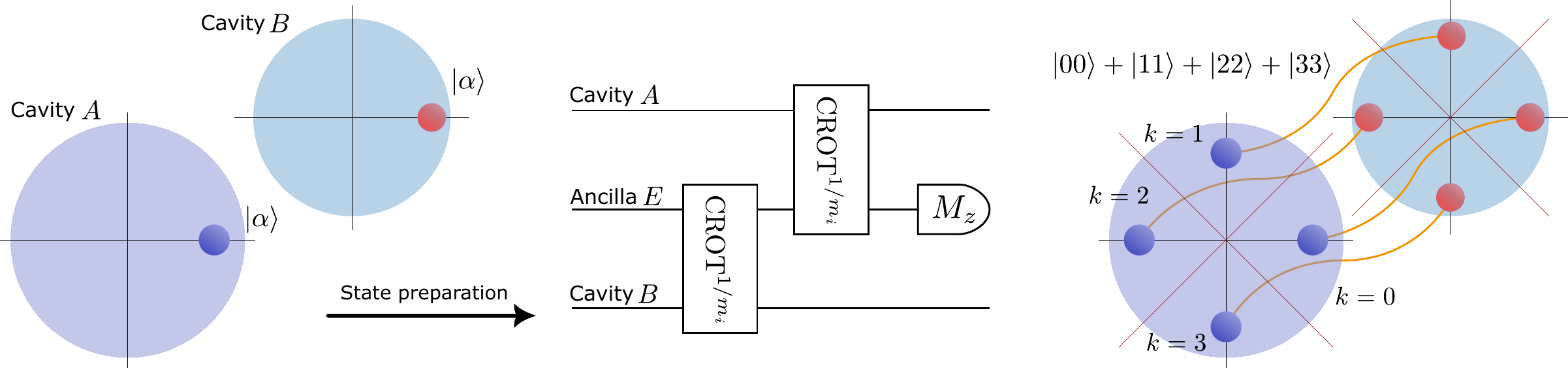}%
        \end{minipage}
        \caption{Schematic representation of the entangled state preparation process. We start with two non-entangled cavities, each prepared in an initial primitive state $\ket{\Theta}$, here represented  by a coherent state $\ket{\alpha}$. The state preparation includes the introduction of an ancilla cavity $E$ and the application of two $\CROT$ gates. The ancilla is then measured, leaving us with an entangled two-cavity state depicted on the right. The entanglement entropy of the final state is $E_i = \log{ \qty(m_i)}$. In this example, we use $m_i=d=4$.}
        \label{fig: state preparation}
    \end{figure*}

    \subsection{Aborting the distillation protocol} \label{sec: Aborting and restarting the protocol}

        To enhance the distillation fidelity, the parties restart the protocol based on the expected fidelity conditioned on the measurement outcomes $\mathcal{F}\qty(A_1,B_1,A_2,B_2)$. This approach requires two-way communication instead of one-way communication and comes at the cost of a reduced success probability $1-p_{\text{abort}}$.

        In Figure \ref{fig: aborting}, we compare the average infidelity of the distilled state under different abort criteria $\mathcal{F}\qty(A_1,B_1,A_2,B_2)\leq f_\mathrm{cut}$ for $f_\mathrm{cut}=0.9$ and $f_\mathrm{cut}=0$, the latter indicating that the protocol is never aborted. 
        The option to abort becomes particularly effective when the qudit dimension $d$ is increased while keeping $m_i$ fixed. By partitioning the cavity phase space into more sections, we effectively increase the measurement resolution, enabling us to more accurately determine when to abort the protocol, thereby improving performance. Therefore, we do not assume $d=m_i$ in Figure \ref{fig: aborting}.
        The probability of aborting the protocol is also shown, emphasizing the trade-off between the achievable fidelity of the distilled state and the probability of successful distillation.

        A notable feature is the non-monotonic behavior of the infidelity as a function of the error rate $\gamma$.
        This behavior has two sources.
        First, the abort criterion is based on the outcome-conditioned expected fidelity $\mathcal{F}\qty(A_1,B_1,A_2,B_2)$.
        As the physical error rate decreases, some outcome classes that were previously rejected can cross the cutoff $f_\mathrm{cut}$ and become accepted.
        These newly accepted outcomes can have relatively low fidelity compared with the already accepted ensemble, thereby increasing the conditional average infidelity of the non-aborted runs.
        Second, because $m_c$ is optimized separately at each error rate, the optimal value of $m_c$ can change as $\gamma$ varies, producing additional non-monotonic features in both the infidelity and the abort probability.
        We include curves for all different $m_c$ values in Appendix \ref{sec: all m_c} for completeness.

\section{Implementing the protocol using Bosonic systems} \label{sec: bosonic-main}

The primary application discussed in this paper targets superconducting circuit-QED hardware, where entanglement distillation involves the transfer of quantum information between two cavities associated with distinct cryogenic modules interconnected via microwave links.
Thus, the precise description of the protocol requires translating the discrete qudit description used in the previous sections into a continuous-variable bosonic description, which we present in this section.


\subsection{Preparing the initial state} \label{section: Creating the initial state}

    We now detail the procedure for initializing the entangled qudit state given in Eq. \eqref{eq: initial state} in a two-cavity system. 
    This involves translating the discrete qudit description used thus far into bosonic states described by continuous variables.
    Our starting point is a product state of the two cavities, each prepared in a primitive state $\ket{\Theta}$ that is approximately localized in one angular section of the cavity phase space. 

   The qudit basis can be constructed from rotated primitive states 
    \begin{align}
        \ket{k^{\text{CV}}}\equiv R_{2 k\pi/{m_i}} \ket{\Theta} \;,
    \end{align}
    where $R_\theta \equiv e^{i\theta \hat{n}}$ is the phase-space rotation operator with $\theta$ the rotation angle 
    and $\hat{n}$ the number operator. 
       
    To create the highly entangled initial state with entanglement entropy $\log_2 m_i$, we use a modified version of the  controlled rotation ($\text{CROT}$) gate introduced in Ref. \cite{Grimsmo_2020},

    \[
        \CROT \coloneqq e^{i\frac{2\pi}{m_i} \hat{n}_A \hat{n}_B} \;,
    \]
    with $\hat{n}_{A\qty(B)}$ the number operator of the mode in cavity $A\qty(B)$.

    This gate can be implemented by applying a coupling Hamiltonian $\hat{n}_A \hat{n}_B$ for a duration of $\frac{2 \pi}{m_i}$. Counterintuitively,  the entanglement generated by the gate increases as the gate duration decreases. Furthermore, this shorter duration can also reduce the infidelity of the entangled state.
    
    The initialization protocol uses an ancilla cavity $E$ in addition to cavities $A$ and $B$ (Fig. \ref{fig: state preparation}). We initialize all three cavities in $\ket{0^{\text{CV}}}$ and apply two $\CROT$ gates, once between cavities $A$ and $E$ and once between cavities $B$ and $E$. We then measure the ancilla in its logical $Z$ basis, and register the result as $x$. 
    
    We choose the primitive state such that the states $\ket{k^\text{CV}}$ are approximately orthogonal. Under this condition, the procedure creates a channel $\mathcal{C}_x$ that acts on the initial state as (Appendix \ref{sec: bosonic operations})
    \begin{multline} \label{eq: creation channel}
        \mathcal{C}_x \qty(\ket{0^{\text{CV}}}_A\ket{0^{\text{CV}}}_B\ket{0^{\text{CV}}}_E)
        \\
        = \frac{1}{\sqrt{m_i}} \sum_{k=0}^{m_i-1} e^{-2ikx\pi/m_i} \ket*{ k^{\text{CV}}}_A\ket*{ k^{\text{CV}}}_B 
        \\
        = Z_A^{-x} \frac{1}{\sqrt{m_i}} \sum_{k=0}^{m_i-1} \ket*{ k^{\text{CV}}}_A\ket*{ k^{\text{CV}}}_B \;.
    \end{multline}
    
    Finally, applying a local unitary  $Z_A^{x}$ leaves us with the desired initial entangled state $\ket{\psi_\Theta} = \frac{1}{\sqrt{m_i}} \sum_{k=0}^{m_i-1} \ket*{ k^{\text{CV}}}_A\ket*{ k^{\text{CV}}}_B$.
    While $Z_A$ corresponds to photon addition, which is challenging to realize \cite{gertler2021protecting}, this operation can also be implemented by updating the obtained initial state in software.

    A similar entangled state can be prepared 
    using only a single $\mathrm{CROT}^{1/m_i}$ gate between Alice's and Bob's cavities, without the need for an ancilla cavity. The resulting state $\frac{1}{\sqrt{m_i}} \sum_{k=0}^{m_i-1} \ket{k^\text{CV}}_A \ket{k_+^\text{CV}}_B$ is similar to $\ket{\psi_\Theta}$, but with one of the cavities transformed to the dual basis. This state can be distilled using a similar protocol, where one of the parties acts in the dual basis while the other acts in the computational basis during the first stage, and vice-versa during the second stage.

    \subsection{Continuous-variable description of the distillation protocol} \label{subsection: Continuous vs. discrete and the effect of squeezing}

    To describe the protocol in a continuous-variable framework, we proceed to redefine the operations introduced in Section \ref{sec: Entanglement distillation protocol} as completely positive maps from the bosonic space onto itself.
    
    We first define the angular section (logical $Z$) measurement as a set of POVMs \cite{Holevo_2011, Grimsmo_2020}
     \begin{align} \label{eq: bosonic_POVM}
        M_k = \frac{1}{\pi}\sum_{m,n=0}^\infty \Gamma_{m,n}^k \ket{m}\bra{n} \;,
    \end{align}
    where $k\in\qty{0,...,d-1}$ is the outcome corresponding to the $k^{\text{th}}$ angular section, $\ket{m}$ and $\ket{n}$ are the Fock-space basis states, and $\Gamma_{m,n}^k$ are coefficients described in Appendix \ref{sec: bosonic operations}. Here, $d$ determines the phase measurement resolution.

    The $\Delta_c$-modular phase measurement with outcome $x\in\qty{0,...,\Delta_c-1}$ is defined analogously to Eq. \eqref{eq: modular phase measurement} as
    \begin{align} \label{eq: bosonic modular phase measurement}
    P_x^{\Delta_c} = \sum_{n=0}^{\mathclap{m_c-1}} M_{n\Delta_c + x} \;,
    \end{align}
    with $m_c=d/\Delta_c$.

    This modular phase measurement can be implemented by applying an interaction Hamiltonian $a^{m_c}r^{\dag} + \text{h.c.}$, where $a$ and $r$ are the annihilation operators of the measured mode and a buffer mode, respectively. The phase obtained from an adaptive homodyne measurement~\cite{Wiseman_1998ahd, Martin_2020ahd_experiment, Hillmann_2022arne_noisy_measurements, berry2002adaptivephasemeasurements} of the field leaking from the buffer mode corresponds to the phase of the cavity state modulo $2\pi / m_c$. This procedure may be accompanied by autonomous stabilization techniques to maintain the average photon number in the measured mode throughout the measurement process \cite{grimm2020stabilization}.

    The completely positive map that corresponds to the parties measuring the results $A_1$ and $B_1$, is $\mathcal{P}^{A_1,B_1} \qty( \bullet ) = P_{A_1}^{\Delta_c} \otimes P_{B_1}^{\Delta_c} \bullet P_{A_1}^{\Delta_c} \otimes P_{B_1}^{\Delta_c} $. 
    To correct the dephasing errors, Alice and Bob rotate their cavities by $-A_1 2 \pi / d$ and $\qty(-B_1 + u_B \Delta_i) 2 \pi / d$ (see Protocol \ref{pro: general protocol}), corresponding  to the unitary map 
    \begin{multline}
        \mathcal{U}_{A_1,B_1}^\phi \qty(\bullet) = 
        \\
        R_{-A_1 2 \pi / d}^A \otimes R_{\qty(-B_1 + u_B) 2 \pi / d}^B 
        \bullet 
        \\
        R_{-A_1 2 \pi / d}^{A\dagger} \otimes R_{\qty(-B_1 + u_B) 2 \pi / d}^{B\dagger} \;.
    \end{multline}

    The $\Delta_f$-modular measurements of the photon number (see Eq. \eqref{eq: dual-basis modular measurement}), which correspond to measuring the qudit in its dual basis, can be expressed as
    \begin{align}
        P_{x+}^{\Delta_f} = \sum_{j=0}^{\infty} \ket{x + j \Delta_f }\bra{x + j \Delta_f} \;,
    \end{align}
    where $\ket{x + j \Delta_f}$ are the Fock-space basis states, and $x\in\qty{0,...,\Delta_f-1}$ is the measurement outcome. 
    Methods for implementing modular photon number measurements are described in Refs. \cite{Grimsmo_2020,Sun_2014, wang2020mod_ph_number}.
    The map corresponding to the parties obtaining outcomes $A_2$ and $B_2$ is $\mathcal{P}^{A_2,B_2}_+ \qty( \bullet ) = P_{A_2+}^{\Delta_f} \otimes P_{B_2+}^{\Delta_f} \bullet P_{A_2+}^{\Delta_f} \otimes P_{B_2+}^{\Delta_f} $. 
    
    Compensating for photon loss errors by adding photons to a cavity is a non-trivial operation \cite{gertler2021protecting}. If this operation is omitted, the resulting entangled state may differ from the original target state but remains a valid entangled resource. 
    Therefore, it is more practical to handle photon loss in software by updating the expected resource state based on measurement outcomes and using this updated state in subsequent steps. Consequently, when evaluating the fidelity of the distilled state, we adjust the target state to match the actual state produced.
     
    When applying the concatenation of all the aforementioned channels to the initial state $\ket{\psi_\Theta}$ (subsection \ref{section: Creating the initial state}), the distillation protocol results in the final state

    \begin{align}
        \rho = ~ \sum_{\mathclap{A_1,A_2,B_1,B_2}} ~ 
        \mathcal{P}^{A_2, B_2}_+ \circ \mathcal{U}_{\phi}^{A_1,B_1} \circ \mathcal{P}^{A_1, B_1} \circ \mathcal{N} \qty(\todm{\psi_\Theta}) \;,
    \end{align}
    where the sum runs over all possible measurement outcomes and $\mathcal{N}$ represents the loss-dephasing channel (see Appendix \ref{sec: The loss-dephasing channel}).


    \subsection{Performance of the protocol under continuous descriptions}

        \begin{figure}[t] 
    \begin{minipage}{\dimexpr\linewidth\relax}           \includegraphics[width=\linewidth]{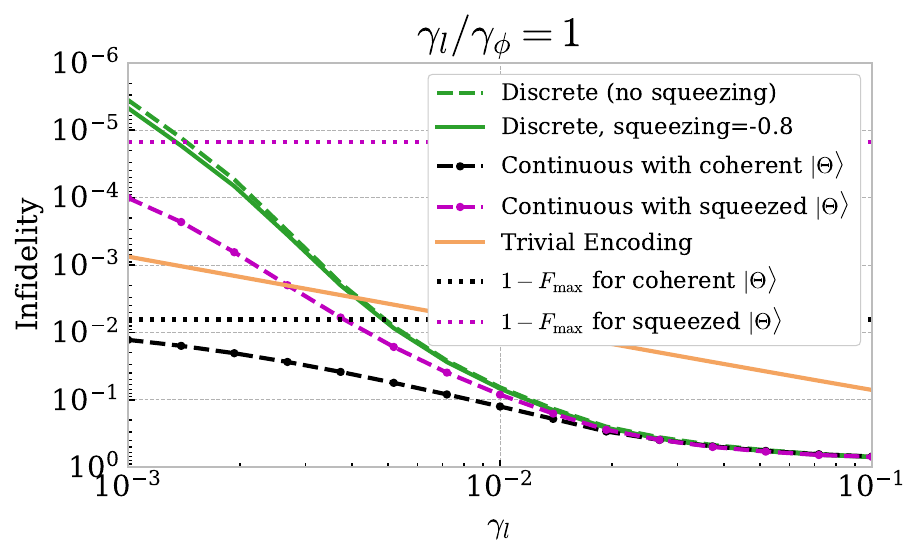}    
        \end{minipage}
        \caption{Simulated infidelity of the final distilled state as a function of the loss and dephasing rates with $\gamma_l=\gamma_\phi=\gamma$. We compare the discrete model with $\bar{n}=49$ (green lines) to the continuous-variable model for a coherent primitive state $\ket{\Theta}=\ket{\alpha=7}$ (dashed gray line) and a squeezed primitive state $\ket{\Theta}=\ket{\alpha=7, \zeta = -0.8}$ (dashed magenta line), with $\alpha$ and $\zeta$ the amplitude and squeezing parameters, respectively.
        Dotted lines represent the limit given by Eq.~\eqref{eq:continuous-limit} for each corresponding curve.
        The discrete model used here is improved to account for the full photon-number distribution of the squeezed state, as described in the Appendix section~\ref{subsection: improved_loss_model}. 
        The dashed (solid) green line shows the discrete model infidelity for a distribution of a squeezed state with $\zeta=-0.8(0)$. 
        The infidelity of the trivial encoding, shown in orange, serves as a baseline for comparison.
        For all cases, we use $m_i=d=16,m_c=8,m_f=2$. 
        }
        \label{fig: squeezing effect}
    \end{figure}

    The discrete qudit model outlined in section \ref{sec: discrete error model} provides a convenient framework for describing the distillation protocol. However, it overlooks certain features of bosonic systems, such as the nonzero overlap between primitive states associated with different angular sections. The overlap between a primitive state $\ket{\Theta}$ and the $k^\mathrm{th}$ section can be quantified as
      \begin{align} \label{eq: delta overlap}
        \delta_k \equiv \bra{\Theta} {M_{k} \ket{\Theta}} \;,
    \end{align}
    where $M_k$ is the POVM defined in Eq. \eqref{eq: bosonic_POVM}.

    Therefore, even in the absence of loss-dephasing noise, the state can be detected in sections other than the initial one, which can be interpreted as a result of quantum noise.
    
    In the first stage of the protocol, a dephasing error of more than $\left\lceil{\Delta_c/2}\right\rceil$ sections in either direction results in an unsuccessful correction. We can use this understanding to bound the protocol infidelity due to the overlap with other sections by 
    \begin{align} \label{eq:continuous-limit}
        1-F_\text{max} = 2 \delta_{\left\lceil{\Delta_c/2}\right\rceil} \;.
    \end{align}
    This limitation of the protocol can be mitigated for a coherent primitive state by increasing its amplitude or by squeezing it along the radial axis, thereby reducing the overlap with adjacent sections. However, this comes at the expense of increased photon loss.

\begin{figure}[t]
        \centering
        \includegraphics[width=\linewidth]{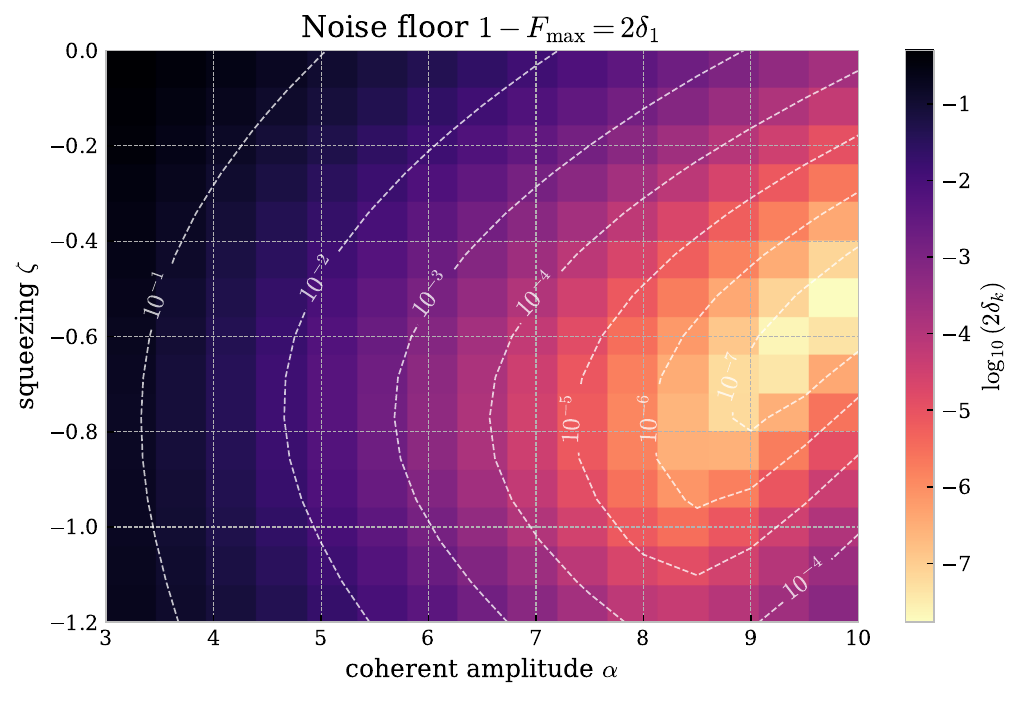}
        \caption{Quantum-noise floor of the continuous-variable protocol as a function of the coherent amplitude $\alpha$ and squeezing parameter $\zeta$. The plotted quantity is the bound $1-F_\text{max}=2\delta_1$, corresponding to Eq.~\eqref{eq:continuous-limit} for $d=m_i=16$ and $m_c=8$. Colors show $\log_{10}(2\delta_1)$, and dashed contours indicate constant values of the noise floor. Increasing $\alpha$ and applying radial squeezing reduce the floor by improving localization within each angular section. There is an optimal value of $\zeta$ for each $\alpha$, as excessive squeezing can increase the overlap with adjacent sections near the origin.}
        \label{fig: quantum-noise-floor}
    \end{figure}

    Figure~\ref{fig: quantum-noise-floor} shows the corresponding quantum-noise floor as a function of the coherent amplitude $\alpha$ and squeezing parameter $\zeta$, for the same protocol parameters as Fig.~\ref{fig: squeezing effect}. For $d=m_i=16$ and $m_c=8$, we have $\Delta_c=2$, so Eq.~\eqref{eq:continuous-limit} gives $1-F_\text{max}=2\delta_1$.

    Another approximation made in the discrete model is a heuristic photon-loss noise model (Appendix \ref{sec: discrete errors}) that uses the average photon number $\bar{n}$ without considering the photon number distribution. This distinction is particularly relevant for primitive states with large photon-number uncertainty, such as squeezed states.
    
    To evaluate the accuracy of the discrete model, we compare the protocol performance using this model to the numerically simulated infidelity obtained with the continuous-variable model for coherent and squeezed primitive states (see Fig. \ref{fig: squeezing effect}). 
    To evaluate the distillation infidelity we use a decoding map $\mathcal{D}$ from the continuous-variable bosonic Hilbert space to a $d$-dimensional logical Hilbert space (Appendix \ref{sec: decoding map}).
    
    So far, in the discrete model, we have used the average photon number $\bar{n}$ to determine the loss rate. However, this approach does not account for the photon number distribution, which is particularly relevant for primitive states with large photon-number uncertainty, such as squeezed states. To address this, we improve the discrete model by considering the full photon-number distribution of the squeezed state (Appendix \ref{subsection: improved_loss_model}). The results of this improved model are also shown in Fig.~\ref{fig: squeezing effect}.
    
    The numerical results indicate that the discrete model provides a good approximation when the protocol infidelity exceeds the error introduced by quantum noise. For sufficiently low loss and dephasing error rates, the continuous-variable model predicts that the fidelity of the distilled state saturates to the quantum noise limit. Reducing the effect of quantum noise by squeezing a coherent primitive state significantly enhances the achievable distillation fidelity.

    The results show that the protocol becomes increasingly effective as the physical error rate decreases, suggesting the presence of a pseudo-threshold, i.e., a break-even point below which it becomes advantageous to apply the protocol rather than rely on trivial encoding. 
    For the squeezed coherent primitive state with the parameters considered in Fig.~\ref{fig: squeezing effect}, this break-even point is $\gamma_b = 2.8 \times 10^{-3}$, within roughly one order of magnitude of error rates reported for short-distance microwave interconnects between cryogenic superconducting modules, $\gamma \sim 10^{-2}$~\cite{niu2023linkent, song2024linkentangler, Mollenhauer_2025linkent}. 
    Appendix~\ref{sec: unequal-loss-dephasing} shows the corresponding break-even behavior for unequal loss and dephasing rates.


    We can also benchmark against a multi-copy qubit scheme by matching the average excitation number. Each qubit in a Bell pair is excited with probability $1/2$, so a cavity with $\bar n=49$ photons carries the same average excitation as a 98-qubit register. The parameters of Fig.~\ref{fig: squeezing effect} thus correspond to two such registers, i.e. on the order of 98 noisy Bell pairs. This comparison is unfavorable to the bosonic distillation scheme: the two Bell-pair registers span an exponentially larger Hilbert space and could in principle distill a far higher-fidelity pair from those many copies. The advantage of the bosonic approach lies instead in hardware efficiency: the high-dimensional resource is stored and manipulated using only two bosonic modes and their control hardware, rather than the $\sim200$ physical qubits the excitation-matched comparison would require.

    \subsection{Other Noise Sources}

    \begin{figure}
        \centering
        \includegraphics[width=\linewidth]{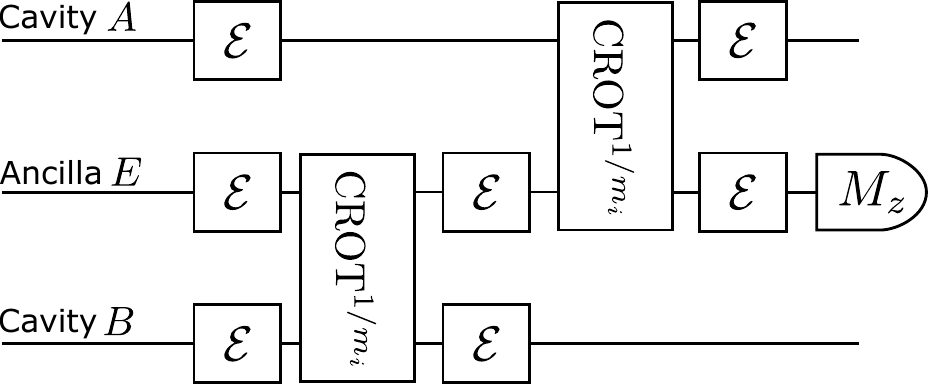}
        \caption{All possible error locations in the initialization process.}
        \label{fig:init-errors}
    \end{figure}

    Throughout the paper, we have only considered the effects of loss and dephasing errors in the transmission process. In this section, we discuss the effects of other realistic noise sources.

    \emph{Initialization errors} -- First, we note that the initialization process can be affected by errors, as shown in Fig. \ref{fig:init-errors}. 
    For dephasing and loss errors, these errors can be propagated to the end of the initialization process, and can be absorbed into the loss and dephasing errors in the transmission process. 

    The propagation is done by the $\CROT$ gate through the following relation~\cite{Grimsmo_2020}:
    \begin{align} \label{eq:CROT_noise}
        \CROT \hat{L}_k^a R_{\theta}^a = \hat{L}_k^a R_{\theta}^a R_{\theta_k}^b \CROT 
    \end{align}

    with $\theta_k = \frac{2\pi k}{m_i}$, and $\hat{L}_k^{a(b)}$ is the loss of $k$ photons on cavity $a(b)$.

    Using this relation, we can propagate the loss and dephasing errors to the end of the initialization process. For the ancilla cavity, which is measured at the $Z$ basis, the dephasing errors do not affect the measurement outcome. However, loss errors on the ancilla cavity can affect the measurement outcome, and will result in a wrong correction on the initial state. This can be treated as a loss error on the initial state.

    \begin{figure}
        \centering
        \includegraphics[width=\linewidth]{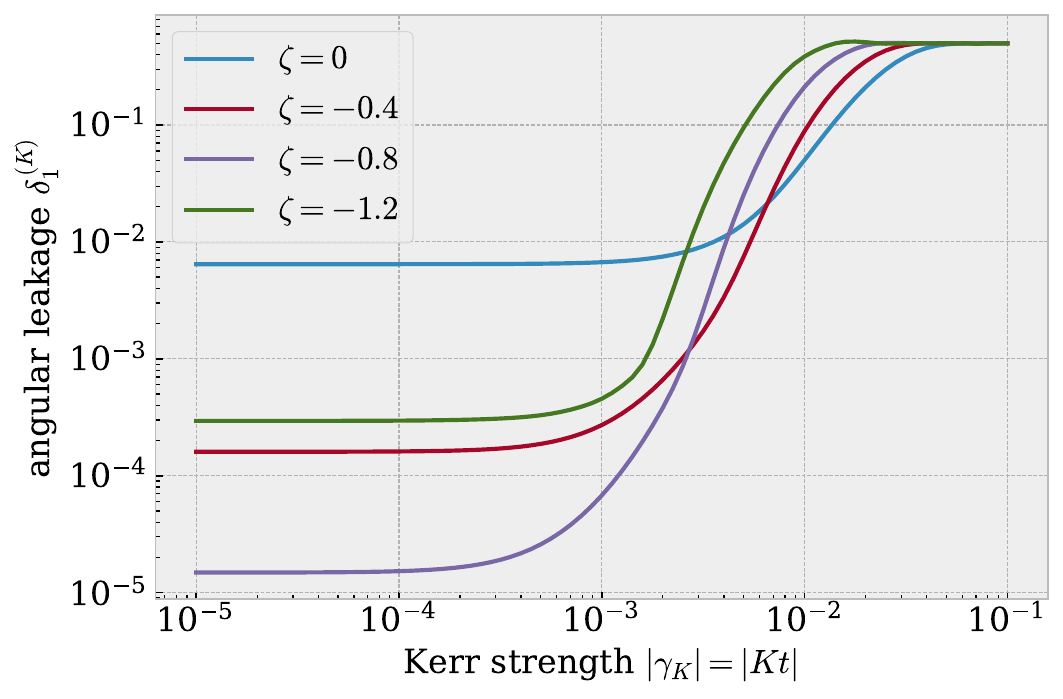}
        \caption{Quantum noise floor of the initial state with Kerr non-linearity as a function of the Kerr strength $\gamma_K=Kt$ for different squeezing parameters $\zeta$. The initial state is a squeezed coherent state with $\alpha=7$ and $d=m_i=16$ and we use $m_c=8$. The Kerr nonlinearity shears the state in phase space, which can increase the overlap with adjacent angular sections. The effect of this error is more pronounced for strongly squeezed states, which have broader photon-number distributions.}
        \label{fig:kerr_leakage}
    \end{figure}

    \emph{Kerr nonlinearity} -- Another source of errors is the Kerr non-linearity~\cite{Kirchmair_2013_Kerr}.

    This is the application of the Hamiltonian $H_K = \frac{K}{2} a^\dagger a^\dagger a a$ on the cavity, where $K$ is the Kerr nonlinearity strength.
    Since this error does not commute with the loss errors, we will only consider the effect of this error on the initial state in order to get a sense of its impact, and further analysis is required to analyze the full effect of this error on the protocol.

    Defining \(\gamma_K=Kt\), the unitary generated by \(H_K\) is
\begin{align}
    U_K(t)
    &= \exp\left[-\frac{i\gamma_K}{2}n(n-1)\right] \nonumber\\
    &= e^{i\gamma_K\bar{n}^2/2}
       e^{-i\gamma_K(2\bar{n}-1)n/2}
       e^{-i\gamma_K(n-\bar{n})^2/2},
\end{align}
where
\begin{align}
    \bar n
    = \langle\Theta_{\alpha,\zeta}|n|
      \Theta_{\alpha,\zeta}\rangle
\end{align}
is the mean photon number of the primitive state. The first factor contributes only a global phase, while the second produces a deterministic phase-space rotation. The latter can be compensated by applying the inverse rotation, or equivalently by updating the phase reference of the cavity. The remaining, intrinsically nonlinear contribution is therefore
\begin{align}
    U_{\mathrm{nl}}(\gamma_K)
    = \exp\left[-\frac{i\gamma_K}{2}
      (n-\bar n)^2\right].
\end{align}
This term applies a photon-number-dependent phase and consequently shears the state in phase space.

    To quantify this effect, we apply \(U_{\mathrm{nl}}\) to a primitive state and calculate its overlap with the angular-section POVM elements \(M_k\):
    \begin{align}
        \delta_k^{(K)}(\alpha,\zeta,\gamma_K)
        =
        \langle\Theta_{\alpha,\zeta}|
        U_{\mathrm{nl}}^\dagger(\gamma_K)
        M_k
        U_{\mathrm{nl}}(\gamma_K)
        |\Theta_{\alpha,\zeta}\rangle .
    \end{align}

    Figure~\ref{fig:kerr_leakage} shows \(\delta_1^{(K)}\) as a function of the dimensionless Kerr strength \(|\gamma_K|=|Kt|\) for \(\alpha=7\) and several squeezing strengths. In the weak-Kerr regime, the curves approach their intrinsic, zero-Kerr leakage floors, with moderate phase squeezing---in particular, \(\zeta=-0.8\) among the values shown---providing the strongest suppression. As \(|\gamma_K|\) increases, the more strongly squeezed states become sensitive to Kerr evolution at smaller values of \(|\gamma_K|\). 

    We can see that for $\zeta = -0.8$, the leakage increase becomes dominant roughly at $|\gamma_K|=2 \times 10^{-4}$. Since recent works have demonstrated that the Kerr nonlinearity in superconducting cavities can be as low as $K/2\pi \sim 1$ Hz~\cite{Milul}, this suggests that the Kerr nonlinearity can be neglected for a protocol with a duration of less than $t\sim 200$ $\mu$s.


    \begin{figure}
        \centering
        \includegraphics[width=\linewidth]{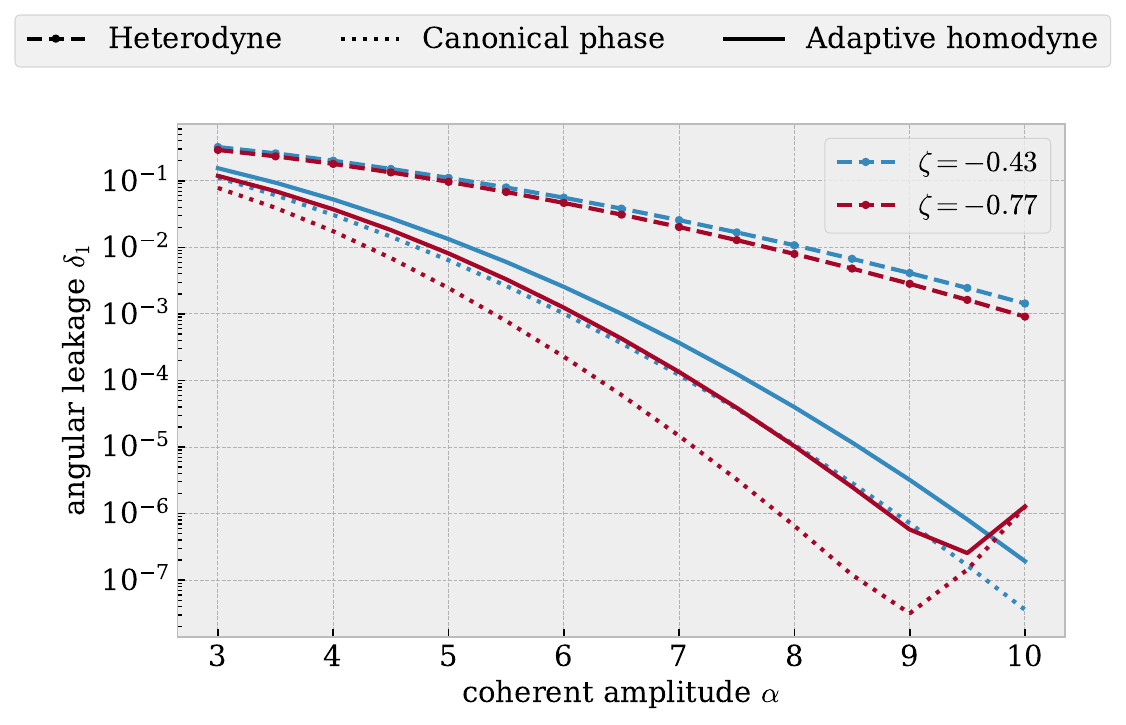}
        \caption{Overlap of the primitive state with the POVM elements corresponding to different phase measurements. The canonical phase measurement (dotted) is the optimal phase measurement used in Fig.~\ref{fig: squeezing effect}. Heterodyne measurements (dashed) offer much worse performance, and thus adaptive homodyne measurement (solid) is required to achieve satisfactory performance.}
        \label{fig:phase_measurement}

    \end{figure}




    \emph{Measurement imperfections} -- Finally, the protocol ideally assumes a sharp canonical phase measurement that perfectly resolves the prescribed angular sectors. In practice, the phase must be inferred using a physical receiver, such as heterodyne detection or adaptive homodyne detection~\cite{Wiseman_1998ahd, Hillmann_2022arne_noisy_measurements}.

    Here, we give an estimation of the effect of imperfect measurements by looking at the overlap of the primitive state with the POVM elements of the measurement.
    The corresponding angular leakage is then
    \begin{align}
        \delta_1^{(x)}
        =
        \mel{\Theta_{\alpha,\zeta}}{M_1^{(x)}}{\Theta_{\alpha,\zeta}},
    \end{align}
    where $M_1^{(x)}$ is the coarse-grained POVM element for measurement scheme $x\in \{\mathrm{canonical}, \mathrm{heterodyne}, \mathrm{adaptive}\}$.

    Figure~\ref{fig:phase_measurement} shows $\delta_1^{(x)}$ as a function of the coherent amplitude $\alpha$ for fixed squeezing parameters $\zeta\in\{-0.43,-0.77\}$. Heterodyne detection produces a substantial increase in angular leakage relative to the ideal canonical phase measurement, in some regimes by several orders of magnitude. By contrast, adaptive homodyne detection remains much closer to the canonical limit and introduces only a modest increase in $\delta_1$. This comparison indicates that adaptive phase estimation may preserve much of the advantage of the ideal phase measurement while providing a more physically accessible implementation.

    \section{Downstream Applications}

    \begin{figure*}[t]
        \centering
        \includegraphics[width=0.7\linewidth]{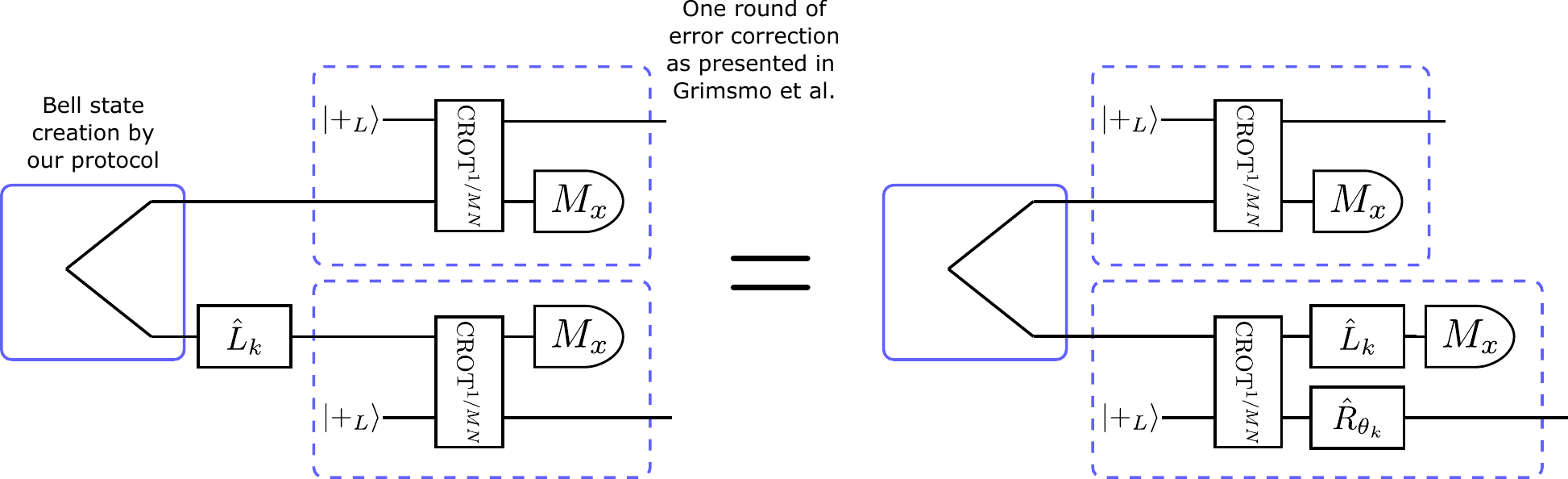}
        \caption{Photon-addition corrections can be tracked as software frame updates during subsequent operations. We illustrate this for a subsequent error-correction round based on the architecture of Grimsmo et al.~\cite{Grimsmo_2020}. A loss event $\hat{L}_k$ before the $\text{CROT}^{1/MN}$ gate is equivalent to the same loss event after the gate together with a phase rotation $R_{\theta_k}$, where $\theta_k=\frac{\pi k}{MN}$.
        The loss event is absorbed in the subsequent measurement because it is a phase measurement, and the phase rotation on the outgoing qubit can be corrected virtually by updating the phase reference of the cavity. Here, $M$ and $N$ are the orders of the rotation-symmetric codes as defined in Ref.~\cite{Grimsmo_2020}.}
        \label{fig: consumption}
    \end{figure*}

    A natural downstream use of the distilled state is as a link-layer resource for modular, networked quantum computation. In the architecture of Grimsmo et al.~\cite{Grimsmo_2020}, logical qubits are encoded in rotation-symmetric bosonic modes, and computation can be performed using code-native gates together with teleportation-based error correction. Our protocol is compatible with this setting: for $m_f=2$, the distilled state is an encoded bosonic Bell pair between two modules, and the measurement record of the distillation protocol determines the corresponding logical frame.

    This shared Bell pair can be consumed to mediate a non-local operation between bosonic processors. Each module can locally couple its data mode to one half of the distilled pair using controlled-rotation operations, measure the auxiliary modes, and communicate the outcomes. In this way, the distilled pair enables qubit teleportation or non-local parity measurement between encoded qubits.

    For example, this could be used in networked quantum-error-correction proposals, such as distributed hyperbolic Floquet codes~\cite{Sutcliffe_2025_distributed,vaknin2026faulttolerantdistributedquantumcomputing}, where non-local pair measurements are supplied by Bell pairs. In our setting, these Bell resources are generated and purified directly in bosonic hardware, reducing the need for large registers of communication qubits dedicated to conventional multi-copy entanglement purification.

    We can also use this format to explain how to account for photon-loss errors in software. 
    We explain this for the case where the following task is an error correction round~\cite{Grimsmo_2020}, and the same frame-update logic applies to other tasks such as pairwise measurements or teleportation. 
    
    The error correction round is based on a $\text{CROT}^{1/MN}$ gate, where $M$ and $N$ are the orders of the rotation-symmetric codes, followed by an $X$-basis measurement of the distilled qubit. 
    This is presented in Fig. \ref{fig: consumption}.
    
    Given a case where $k$ loss errors have been detected in our protocol, we can account for them in the error correction round. 
    As given in Eq.~\eqref{eq:CROT_noise} and presented in Fig.~\ref{fig: consumption}, the loss of $k$ photons on the control qubit propagates through the $\text{CROT}^{1/MN}$ gate as a phase rotation of $\theta_k=\frac{\pi k}{MN}$ on the target qubit, in addition to the original loss of $k$ photons on the control qubit~\cite{Grimsmo_2020}.
    
    The loss of $k$ photons on the control qubit is absorbed in the subsequent $X$-basis measurement, as it is a phase measurement. The phase rotation $\theta_k$ can be corrected virtually by updating the phase reference of the target cavity.

    Rather than physically restoring the canonical Bell-pair representative, we retain a classical loss/phase frame. For a downstream Grimsmo-style error correction cycle, this frame propagates through the controlled-rotation gate as a known phase rotation of the data cavity, which can be absorbed by updating the cavity phase reference or the interpretation of later measurement outcomes.

    \section{Conclusion}

    We have introduced an entanglement distillation procedure for bosonic systems using states with discrete rotation symmetry in phase space. This choice ensures compatibility with the rotation-symmetric bosonic quantum error correction codes introduced in Ref. \cite{Grimsmo_2020}. In particular, the final distilled state with $m_f=2$ corresponds to an encoded Bell state, which can be used in bosonic fault-tolerant quantum computing schemes \cite{Grimsmo_2020} and facilitates  distributed bosonic quantum computing \cite{jiang2007distributed}. 
    Our protocol becomes advantageous when loss and dephasing errors fall below a pseudo-threshold error of $2.8 \times 10^{-3}$, close to the performance currently achievable with remotely connected modules~\cite{niu2023linkent, song2024linkentangler, Mollenhauer_2025linkent}.
        
    Additionally, our distillation protocol can be extended to other bosonic encodings. For instance, by using initial entangled states with translation symmetry in phase space instead of rotation symmetry, the distillation protocol becomes compatible with GKP (grid) quantum error correcting codes \cite{Gottesman_2001,Campagne_Ibarcq_2020GKPexperiment,Sivak_2023GKP_exp, Leviant_2022}. In this case, angular rotations and photon addition are replaced with more straightforward displacements along the position and momentum coordinates. 
    Since our bosonic distillation protocol is limited to a single-shot application, concatenation with a traditional distillation protocol at the logical level could be used to achieve even higher fidelities.

    In this work, we assumed that the noise channel only impacts the transmission step between Alice and Bob, supplemented by a brief analysis of additional error sources. Future work should address these error sources in greater detail. Following the presented analysis, we expect the distillation procedure to remain robust against these errors.

    Our proposed distillation scheme relies on operations native to bosonic systems, with many of the required building blocks \cite{rojkov2024stabilization, ofek2016parity, Konno_2024, Milul,gertler2021protecting, Hajr_2024} and  quantum links \cite{Wallraff_link, axline2018demand_Schoelkopf, LinkBurkhart.2.030321, almanakly2024deterministicremoteentanglementusing,storz2023loophole} already demonstrated in recent hardware.
    We therefore anticipate that this protocol can be realized using currently available experimental platforms.

    \begin{acknowledgments}

    The authors thank Liang Jiang for his helpful comments on this manuscript. We also thank Peter Leviant for providing us with numerical tools that were used in the simulations of this manuscript.
    All authors acknowledge support from the Marshall and Arlene Bennett Family Research Program.
     SJ and SR acknowledge additional financial support from the Israel Science Foundation ISF Quantum Science and Technologies Grants No. 963/19 and No. 2022/20, and the European Research Council Starting Investigator Grant No. Q-CIRC 101040179. 
     RAF research was also generously supported by the Peter and Patricia Gruber Award and the Koshland Research Fund. 
    SR is the incumbent of the Rabbi Dr Roger Herst Career Development Chair and RAF holds the Daniel E. Koshland Career Development Chair.
     
    \end{acknowledgments}

    \section*{Author Contributions}

    All authors contributed to the development of the ideas and the writing of the manuscript. S.J. performed the numerical simulations and analytical calculations. S.R. and R.A.F. supervised the project.
    The authors acknowledge the use of AI tools for assistance with code generation and writing. All AI-generated use was closely guided. All generated code was checked for correctness.

\section{Data Availability}

The scripts used in this paper and the generated data are available in~\cite{jacoby2026zenodo}.

    \appendix

    \onecolumn

    \section{Error probabilities of the discrete model}

    \subsection{The loss-dephasing channel} \label{sec: The loss-dephasing channel}
    
        The main noise sources in bosonic systems are photon loss and dephasing.  The density matrix $\rho$ of such a system evolves under the combined loss and dephasing channels \cite{Leviant_2022} according to the master equation
        \begin{align}
            \dot{\rho} = \kappa_l \mathcal{D} \qty[\hat{a}] \rho + \kappa_\phi \mathcal{D} \qty[\hat{n}] \rho \;,
        \end{align}
        where $\hat{a}$ and $\hat{n}$ are the annihilation and number operators of the bosonic system,  $\kappa_l$ and $ \kappa_\phi$ are the loss  and dephasing rates, and $\mathcal{D} \qty[\hat{A}]$ is the Lindbladian defined as
        \begin{align}
            \mathcal{D} \qty[\hat{A}]\qty( \bullet ) = \hat{A} \bullet \hat{A}^\dagger - \frac{1}{2}\hat{A}^\dagger \hat{A} \bullet - \frac{1}{2} \bullet \hat{A}^\dagger \hat{A} \;.
        \end{align}
        with $\bullet$ denoting any linear operator over the bosonic system's Hilbert space.

        This allows us to define the loss-dephasing channel as the time evolution superoperator
        \begin{align}
            \mathcal{N} \qty[\kappa_l, \kappa_\phi]= e^{\kappa_l t \mathcal{D}[\hat{a}]+\kappa_\phi t\mathcal{D}[\hat{n}]},
        \end{align}
        with $t$ the evolution time and $\rho(t)= \mathcal{N} \qty[\kappa_l, \kappa_\phi]\left(\rho(0)\right)$. Since the loss and dephasing channels commute, we can decompose the combined channel into two independent channels $\mathcal{N}\qty[\kappa_l, \kappa_\phi ]= \mathcal{N}_L \qty[\gamma_l \equiv 1-e^{-\kappa_l t}] \circ \mathcal{N}_D \qty[\gamma_\phi \equiv \kappa_\phi t]$. 
        These separate channels can be written using the Kraus-operator representation as \cite{Leviant_2022}
        \begin{align} \label{eq: loss channel}
            \mathcal{N}_L \qty[\gamma_l]  & =  \sum_{k=0}^\infty \hat{L}_k \bullet \hat{L}_k^\dagger,\; \hat{L}_k =\sqrt{\frac{\gamma_l^k}{k!}} \qty(1-\gamma_l)^{\hat{n}/2} \hat{a}^k \;
        \\
        \label{eq: dephasing channel}
            \mathcal{N}_D \qty[\gamma_\phi]  & = \frac{1}{\sqrt{2\pi\gamma_\phi}} 
            \int_{-\infty}^\infty
            e^{-\frac{\phi^2}{2\gamma_\phi}} R_\phi \bullet R_\phi^\dagger d\phi \;.
        \end{align}

    \subsection{Discretized errors} \label{sec: discrete errors}

        Using the channels described above, we can construct an approximate error model for the discretized system introduced in Section \ref{sec: discrete error model}.

        The dephasing channel can be interpreted using Eq. \eqref{eq: dephasing channel} as a statistical mixture of phase-space rotations by an angle $\phi$ with a Gaussian probability distribution:
        \begin{align}
            f_D \qty(\phi) =  \frac{1}{\sqrt{2\pi\gamma_\phi}}  e^{-\frac{\phi^2}{2\gamma_\phi}} \;.
        \end{align}
        We discretize the dephasing errors by approximating rotations within the angular interval $\phi \in \qty(\qty(2s-1) \pi / d, \qty(2s+1) \pi / d)$ as a discrete dephasing error by $s$ sections. The corresponding probability for dephasing by $s$ sections is:
        \begin{equation*}
            {p}_s^{D}
            =\int_{\left(2s-1\right)\pi/d}^{\left(2s+1\right)\pi/d}f_D \qty(\phi)d\phi
            =\frac{1}{2}
            \left[\operatorname{erf}\left(\sqrt{\frac{1}{2\gamma_{\phi}}}\left(2s+1\right)\pi/d\right)-\operatorname{erf}\left(\sqrt{\frac{1}{2\gamma_{\phi}}}\left(2s-1\right)\pi/d\right)\right] \;,
        \end{equation*}
        with $\operatorname{erf}$ the error function.

        The probability of losing $l$ photons depends on the photon number distribution of the bosonic state. For approximating the loss probability in the discrete model, we assume that the photon number distribution is sharply peaked around the average photon population $\bar{n}$. Using Eq. \eqref{eq: loss channel}, we can therefore write
        \begin{align}
            \tilde{p}_l^{L} = \langle \hat{L}_l^\dagger \hat{L}_l \rangle \approx \frac{\gamma_l^l}{l!} \qty(1-\gamma_l)^{\bar{n}} \bar{n}^l \;.
        \end{align}

        The sum of the loss probabilities \( p_l^L(\bar{n}) \) does not necessarily equal 1. Thus, we normalize them as follows:
        \begin{align}
                p_l^{L} = \frac{\tilde{p}_l^{L}}{\sum_{j\ge0
                } \tilde{p}_j^{L}} \;.
            \end{align}

\begin{figure} 
            \centering
            \includegraphics[width=0.5\linewidth]{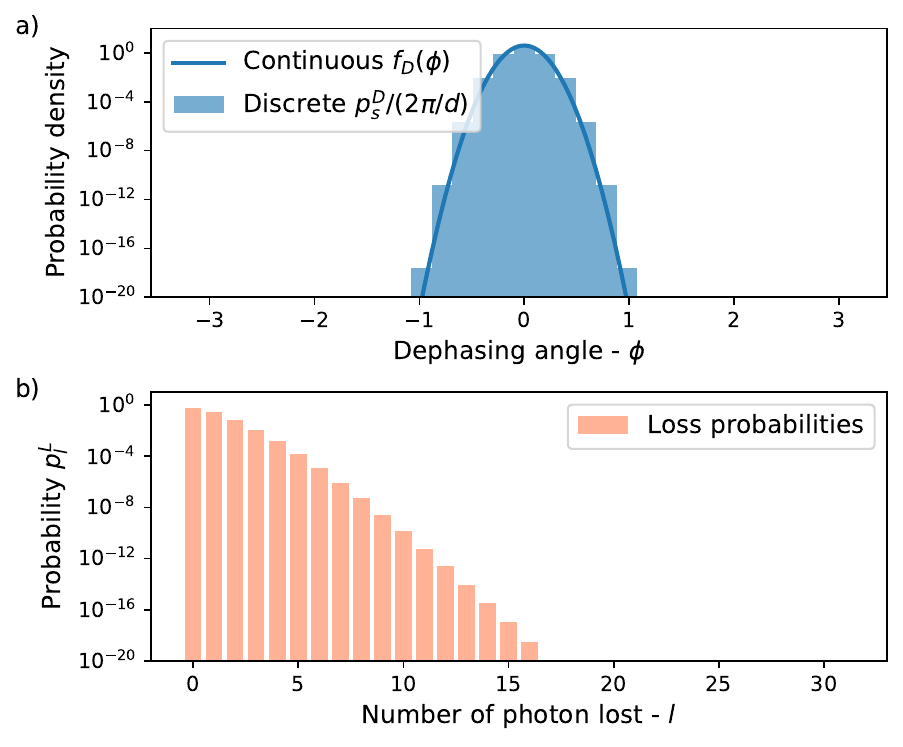}
            \caption{a) Probability density of dephasing by an angle $\phi$ without the discrete approximation ($f_D(\phi)$, solid line) and with the discrete approximation ($p_s^D/(2\pi/d)$, bars), $\gamma_\phi={10}^{-2}$.
            b) Probability of losing $l$ photons with the discrete approximation. Here, $\bar{n}=49$ and $\gamma_l={10}^{-2}$.
            }
            \label{fig:discrete-errors}
        \end{figure}

        We plot the discrete error probabilities in Fig.~\ref{fig:discrete-errors}. This distribution exhibits super-exponential decay, which distinguishes it from typical qubit-based distillation protocols. As a result, we observe that $p_{s=2}^D \ll ({p_{s=1}^{D}})^2$.

    \subsubsection{Discretized loss errors for states with a wide photon number distribution} \label{subsection: improved_loss_model}

        The sharply-peaked approximation can be relaxed by averaging the loss probability over the full photon-number distribution. Let
        \begin{align}
            P(n) = \mel{n}{\rho}{n}
        \end{align}
        denote the probability that the state immediately before the loss channel contains $n$ photons. Starting directly from the loss Kraus operator in Eq.~\eqref{eq: loss channel}, the probability of losing exactly $l$ photons is
        \begin{align}
            p_{l,\mathrm{full}}^{L}
            &= \Tr\!\left(\hat{L}_l \rho \hat{L}_l^\dagger\right)
            = \left\langle\hat{L}_l^\dagger \hat{L}_l\right\rangle \notag\\
            &= \frac{\gamma_l^l}{l!}
            \left\langle
            \hat{n}^{\underline{l}}
            \qty(1-\gamma_l)^{\hat{n}-l}
            \right\rangle \notag\\
            &= \sum_{n=l}^{\infty} P(n) \binom{n}{l}
            \gamma_l^l \qty(1-\gamma_l)^{n-l},
            \label{eq: full-photon-number-loss-probability}
        \end{align}
        where $\hat{n}^{\underline{l}} \equiv \hat{n}(\hat{n}-1)\cdots(\hat{n}-l+1)$ is the falling factorial of the photon-number operator. The final equality follows from
        $\hat{a}^l\ket{n}=\sqrt{n!/(n-l)!}\ket{n-l}$. Equivalently, conditioned on an initial photon number $n$, the number of lost photons follows a binomial distribution, and Eq.~\eqref{eq: full-photon-number-loss-probability} averages this conditional probability over $P(n)$.

        The relation between Eq.~\eqref{eq: full-photon-number-loss-probability} and the sharply-peaked approximation used above is
        \begin{align}
            p_{l,\mathrm{full}}^L
            &= \frac{\gamma_l^l}{l!}
            \left\langle
            \hat{n}^{\underline{l}}
            \qty(1-\gamma_l)^{\hat{n}-l}
            \right\rangle \notag\\
            &= \sum_{n=l}^{\infty} P(n)\binom{n}{l}
            \gamma_l^l\qty(1-\gamma_l)^{n-l} \notag\\
            &\approx \binom{\bar{n}}{l}
            \gamma_l^l\qty(1-\gamma_l)^{\bar{n}-l} \notag\\
            &\approx \frac{\gamma_l^l}{l!}
            \qty(1-\gamma_l)^{\bar{n}}\bar{n}^l
            \equiv \tilde{p}_l^L.
            \label{eq: loss-probability-approximation-chain}
        \end{align}
        The first approximation replaces the photon-number distribution by a value concentrated at $\bar{n}$. The second additionally uses $\bar{n}^{\underline{l}}\approx\bar{n}^l$ and $\qty(1-\gamma_l)^{\bar{n}-l}\approx\qty(1-\gamma_l)^{\bar{n}}$, which are appropriate when $l\ll\bar{n}$ and $l\gamma_l\ll1$.

        In contrast to the approximate weights $\tilde{p}_l^L$, the full-distribution probabilities are normalized without any additional rescaling:
        \begin{align}
            \sum_{l=0}^{\infty} p_{l,\mathrm{full}}^L
            &= \sum_{n=0}^{\infty}P(n)
            \sum_{l=0}^{n}\binom{n}{l}
            \gamma_l^l\qty(1-\gamma_l)^{n-l}
            = \sum_{n=0}^{\infty}P(n)=1.
        \end{align}

        We use this model in Sec.~\ref{subsection: Continuous vs. discrete and the effect of squeezing} to show a better comparison with the continuous model for highly-squeezed states, which have a broad photon-number distribution.

    \section{Analyzing the distillation protocol} \label{sec: analyzing the protocol}

        In order to calculate the fidelity of the distilled state, we follow the system's density matrix from the initial noisy entangled state through the measurements and the corrections by the parties. In the appendices, we do not assume $m_i=d$.

        The initial state is described by Eq. \eqref{eq: initial noisy state} in the main text as follows:
        \begin{align} \label{eq:initial-noisy-state-appndx}
            \rho &= \sum_{\mathclap{\bar{x} \in \mathcal{E}}}
            p\qty(\bar{x})\todm{\varphi_{\bar{x}}} \;,
        \end{align}
        where
            $\ket{\varphi_{\bar{x}}} = 
            L^A_{l_A} L^B_{l_B}
        D^A_{s_A} D^B_{s_B}  \ket{\psi_i}$
        are the error states associated with the error $\bar{x}=\qty(s_A,s_B,l_A,l_B)$. The summation is over all possible errors. Since the initial state $\psi_i$ can be considered as the state corresponding to $\bar{x}=\qty(0,0,0,0)$, we hereafter include it in the summation. Writing the error states explicitly, we have:

        \begin{align} \label{eq: pure noisy state appendix}
            \ket{\varphi_{\bar{x}}} = L^{l_A} L^{l_B} \sum_{k=0}^{m_i-1} \ket{\Delta_i k + s_A}_A \ket{\Delta_i k + s_B}_B \;,
        \end{align}

        with $\Delta_i=d/m_i$.

        In the first distillation stage, which corrects dephasing errors, both parties perform modular-$\Delta_c$ phase measurements, yielding outcomes $A_1$ and $B_1$ for Alice and Bob, respectively. 
        When applying the PVM (Eq. \eqref{eq: modular phase measurement} in the main text)
        \begin{align}
            P_x^{\Delta_c} = \sum_{j=0}^{\mathclap{d/\Delta_c-1}} \todm{ j\Delta_c  + x}
        \end{align}
        to the error state $\ket{\varphi_{\bar{x}}}$, we obtain:
        \begin{multline}
            \ket{\tilde{\varphi}_{\bar{x}}} \equiv P_{A_1}^{\Delta_c} P_{B_1}^{\Delta_c} \ket{\varphi_{\bar{x}}} 
            =
            \\
            \frac{L^{l_A} L^{l_B}}{\sqrt{m_i}} \qty(
            \sum_{n,m=0}^{\mathclap{d/\Delta_c-1}} \ket{\Delta_c n + A_1}_A 
            \bra{\Delta_c n + A_1}_A
            \otimes
            \ket{\Delta_c m + B_1}_B
            \bra{\Delta_c m + B_1}_B
            )
            \sum_{k=0}^{m_i-1} \ket{\Delta_i k + s_A}_A \ket{\Delta_i k + s_B}_B \;.
        \end{multline}

        For the projection to be non-zero, the inner products \(\bra{\Delta_c n + A_1}_A \ket{\Delta_i k + s_A}_A\) and \(\bra{\Delta_c m + B_1}_B \ket{\Delta_i k + s_B}_B\) must be non-zero for certain values of $(n,k,m)$. This occurs only under the constraints:
        \begin{align} \label{eq: first const comp}
            A_1 & = s_A  \pmod{\Delta_i} \;,
            \\ \label{eq: second const comp}
            B_1 & = s_B \pmod{\Delta_i} \;,
            \\
            \label{eq: third const comp}
            A_1 - B_1 & = s_A - s_B \pmod{\Delta_c} \;.
        \end{align}
        resulting in the post-measurement state described by Eq. \eqref{eq: collapsed state}:
        \begin{align} \label{eq: collapsed state appendix}
            \ket{\tilde{\varphi}_{\bar{x}}}
            & = \frac{1}{\sqrt{m_c}} \sum_{k'=0}^{m_c-1} c_{k'} \ket{\Delta_c k' + A_1}_A \ket{\Delta_c k' + \Delta_c u + B_1}_B \;,
        \end{align} 
        with $c_{k'}$ a phase factor resulting from possible loss errors and $u = \frac{\delta_s + A_1-B_1}{\Delta_c}$ an integer.

        Considering the full density matrix again (Eq. \eqref{eq:initial-noisy-state-appndx}), the constraints \eqref{eq: first const comp}--\eqref{eq: third const comp} can also be used to reduce the possible set of errors $\mathcal{E}$, yielding the post-measurement density matrix
        \begin{align}
            \rho_1 & = \sum_{\mathclap{\bar{x} \in \mathcal{E}_1}}
            p\qty(\bar{x})\todm{\tilde{\varphi}_{\bar{x}}} \;,
            \\
            \mathcal{E}_1 & = \qty{\bar{x}=\qty(s_A,s_B,l_A,l_B) \in \mathcal{E} \label{eq: rotation restrictions}
            :
            \begin{array}{lr}
                \scriptstyle s_A = A_1 \pmod{\Delta_i}
                \\
                \scriptstyle s_B = B_1 \pmod{\Delta_i}
                \\
                \scriptstyle s_A - s_B = A_1 - B_1 \pmod{\Delta_c}
            \end{array}
            } \;.
        \end{align}

        We next describe the phase-space rotations that Alice and Bob must apply to realign their states with the $\ket{0}$ state of their respective cavities and to realign the cavities with respect to each other.

        Alice's phase-space rotation is based solely on her measurement outcome $A_1$. She realigns her state with the $\ket{0}$ state by applying $X^{-A_1}$.
        Bob's rotation, on the other hand, depends on both his own result and Alice's. He calculates an integer $u_B$ that maximizes the probability of a successful realignment using a rotation $X^{-B_1+u_B \Delta_c}$.

        After the application of these phase-space rotations, the error state can be written as
        \begin{align} \label{eq: state after first step}
            \ket{\tilde{\varphi}_{\bar{x}}'} = \frac{1}{\sqrt{m_c}}
            L^{l_A} L^{l_B} \sum_{k=0}^{m_c-1} \ket{\Delta_c k }_A \ket{\Delta_c k + s_B - s_A + A_1 - B_1 + u_B \Delta_c }_B 
            \\
            = \frac{1}{\sqrt{m_c}}
            L^{l_A} L^{l_B} \sum_{k=0}^{m_c-1} \ket{\Delta_c k }_A \ket{\Delta_c (k + u + u_B) }_B \;.
        \end{align}

        This equation indicates that the condition for a successful rotation is
        \begin{align} \label{eq: successful rotation condition}
            u=- u_B \pmod{m_c} \;.
        \end{align}
        
        In order to maximize the probability for a successful realignment, we find the value of $u$ that maximizes the likelihood    $\mathcal{L}(u|A_1,B_1)$ \cite{Bravyi_2014}. Since the probability for a specific error $\bar{x}$ is $p\qty(\bar{x}) = p^D_{s_A} p^D_{s_B} p^L_{l_A} p^L_{l_B}$, we can sum over all the different possible errors $\bar{x}$ that result in measurement outcomes $(A_1,B_1)$. The dephasing errors that satisfy the constraints from Eq. \eqref{eq: rotation restrictions} can be expressed as $s_A = A_1 + t \Delta_i$ and $s_B = B_1 + t \Delta_i + u \Delta_c$, where $t,u \in \mathbb{Z}$. Thus, we obtain
        \begin{align} \label{eq:counting_rotation}
            u_B = - \argmax_{u \in \mathbb{Z}} \qty(\sum_{\substack{t=0,\pm1,....}} p^D_{A_1 + t \Delta_i} \cdot p^D_{B_1 + t \Delta_i + u \Delta_c}) \;.
        \end{align}
       The loss probabilities do not affect the calculation and are therefore omitted from this expression.

        The second stage, in which loss errors are corrected, can be better understood by examining the error state from Eq. (\ref{eq: state after first step}) in the dual basis:
        \begin{align*}
        \ket{\tilde{\varphi}_{\bar{x}}'} & = \frac{\hat{P}}{\sqrt{m_c}} \sum_{k=0}^{m_c-1}\ket{\Delta_c k}_A \ket{\Delta_c k}_B 
        = \frac{\hat{P}}{\sqrt{m_c}} \sum_{k=0}^{m_c-1}
        \qty(\frac{1}{\sqrt{d}} \sum_{j=0}^{d-1} e^{-2ij \Delta_c k\pi/d} \cdot \ket{j_+})
        \qty(\frac{1}{\sqrt{d}} \sum_{t=0}^{d-1} e^{-2it \Delta_c k\pi/d} \cdot \ket{t_+})
        \\ & =\frac{\hat{P}}{\sqrt{m_c}d}\sum_{j,t=0}^{d-1} \sum_{k=0}^{m_c-1} e^{{-2i\qty(j+t) \Delta_c k\pi/d}} 
        \ket{j_+}_A \ket{t_+}_B
        = \frac{\hat{P}}{\sqrt{m_c}d}\sum_{j,t=0}^{d-1} \qty(\sum_{k=0}^{m_c-1} e^{{-2i\qty(j+t) k\pi/m_c}})
        \ket{j_+}_A \ket{t_+}_B
        \\ & = \frac{\hat{P}}{\sqrt{m_c}d}\sum_{j,t=0}^{d-1} \qty(m_c \sum_{u=-\infty}^{\infty} \delta_{j,-t + m_c \cdot u})
        \ket{j_+}_A \ket{t_+}_B
        = \frac{\hat{P}}{\sqrt{m_c}\Delta_c}\sum_{t=0}^{d-1} \sum_{u=0}^{\Delta_c}
        \ket{\qty(-t + m_c \cdot u)_+}_A \ket{t_+}_B
        \\ & = \frac{\hat{P}}{\sqrt{m_c\Delta_c}}\sum_{t=0}^{d-1}
        \ket{-t_+^{m_c}}_A \ket{t_+}_B 
        = \frac{\hat{P}}{\sqrt{m_c}}\sum_{t=0}^{m_c-1}
        \ket{-t_+^{m_c}}_A \ket{t_+^{m_c}}_B
        \;.
        \end{align*}
        Here, $\hat{P}=L^A_{l_A} L^B_{l_B} D^B_{\qty(u-u_B) \Delta_c}$ represents the residual error operator. Specifically, the dephasing operator $D^B_{\qty(u-u_B) \Delta_c}$ represents an unsuccessful correction in the first stage. Because any states with $u\neq u_B$ are orthogonal to the desired state -- and thus do not contribute to the fidelity -- we later omit them from the density matrix (without renormalization). Consequently, we can continue analyzing the protocol without including this dephasing operator. Thus, we recover
        Eq. \eqref{eq: state for second step expanded} in the main text:
        \begin{align} \label{eq: state for second step expanded appndx}
            \frac{1}{\sqrt{m_c}}  \sum_{k=0}^{m_c-1} \ket{\qty(k-l_A)_+^{m_c}}_A \ket{\qty(-k-l_B)_+^{m_c}}_B \;.
        \end{align}
        
        The second stage of the protocol starts with both parties performing a modular measurement in the dual basis. Alice and Bob register the results as $A_2$ and $B_2$, respectively. 
        Acting on the error state with the PVMs (first introduced in Eq. \eqref{eq: dual-basis modular measurement} in the main text)
        \begin{align}
        P_{x+}^{\Delta_f} = \sum_{j=0}^{\mathclap{d/\Delta_f-1}} \todmstar{(j\Delta_f + x)_+} \;,
        \end{align}
        
       we can deduce that the state after the dual-basis measurement meets the constraint
        \begin{align}
            l_A+l_B=-\qty(A_2+B_2) \pmod{m_c/m_f} \;.
        \end{align}
        The counterparts to the first two constraints in the computational basis (Eqs. \eqref{eq: first const comp} and \eqref{eq: second const comp}) are trivial. This is because the distance \footnote{Here, “distance” refers to the minimum number of errors required to transform one supporting state into another.} between supporting states (equivalent to $\Delta_i$ in the computational stage)  in Eq. \eqref{eq: state for second step expanded appndx} is $1$.

        This constraint again reduces the possible set of errors, yielding the density matrix
        \begin{align}
            \rho_2 & = \sum_{\mathclap{\bar{x} \in \mathcal{E}_2}}
            p\qty(\bar{x})\todm{\vardbtilde{\varphi}_{\bar{x}}} \;,
            \\
            \mathcal{E}_2 & = \qty{\bar{x}=\qty(s_A,s_B,l_A,l_B) \in \mathcal{E}
            :
            \begin{array}{lr}
                \scriptstyle s_A = A_1 \pmod{\Delta_i}
                \\
                \scriptstyle s_B = B_1 \pmod{\Delta_i}
                \\
                \scriptstyle s_A - s_B = A_1 - B_1 \pmod{\Delta_c}
                \\
                \scriptstyle l_A+l_B=-\qty(A_2+B_2) \pmod{m_c/m_f}
            \end{array}
            } \;,
        \end{align}
        with the state after the second measurement being
        \begin{align} \label{eq:state-after-second-collapse-appndx}
            \ket{\vardbtilde{\varphi}_{\bar{x}}} = 
            \sum_{k=0}^{m_f-1} \ket{\qty( \Delta_f k + A_2)^{m_c}_+ }_A \ket{\qty(-\Delta_f k + B_2 + \Delta_f v)^{m_c}_+ }_B    \;,
        \end{align}
        where $\Delta_f = \frac{m_c}{m_f}$ and $\Delta_f v=-\qty(l_A+l_B+A_2+B_2)$.

        Alice applies a rotation in the dual basis by $-A_2$, which corresponds to the addition or removal of photons. Bob calculates the best possible dual-basis rotation, chosen to maximize the likelihood of successful realignment (similarly to the first stage):
        \begin{align} \label{eq:counting_loss}
            v_B = \argmax_{v \in \qty{0,...,m_f-1}} \qty(
            \sum_{\substack{t=0,\pm1,..., \pm m_c-1
                            \\ j = 0, \pm1, ..., \pm \qty(\Delta_c-1)
                            }} 
            p_{l=- A_2 + t} \cdot p_{l = -B_2 - t - v \frac{m_c}{m_f} + j \cdot m_c}) \;.
        \end{align}

        Bob rotates his cavity in the dual basis by $-v_B\Delta_f - B_2$, resulting in the state
        \begin{align}
            \sum_{k=0}^{m_f-1} \ket{\qty(\Delta_f k)^{m_c}_+ }_A \ket{\qty(-\Delta_f k + \Delta_f \qty(v-v_B))^{m_c}_+ }_B \;.
        \end{align}
       This equation shows that a successful realignment is achieved when
        \begin{align} \label{eq: successful rotation condition}
            v= v_B \pmod{m_c} \;.
        \end{align}
        
        To calculate the fidelity of this distilled state to the desired target state, we can sum over the probabilities for all errors that result in a successful realignment, i.e., for which Eqs. $u_B=-u$, and $v_B=v$ are satisfied. This set is given by
        \begin{align}
            \mathcal{S} = \qty{\bar{x}=\qty(s_A,s_B,l_A,l_B) \in \mathcal{E}
            :
            \begin{array}{lr}
                \scriptstyle s_A = A_1 \pmod{\Delta_i}
                \\
                \scriptstyle s_B = B_1 \pmod{\Delta_i}
                \\
                \scriptstyle s_A - s_B = A_1 - B_1 + u_B \Delta_c \pmod{d}
                \\
                \scriptstyle l_A+l_B=-\qty(A_2+B_2) + v_B \Delta_f \pmod{d}
        \end{array}
        } \;.
        \end{align}

        For specific measurement outcomes $\qty(A_1,B_1,A_2,B_2)$, the fidelity to the target state is the ratio between the total probability of error configurations $\bar{x}$ that lead to a successful realignment and the total probability of the set $\mathcal{E}_2$ of all error configurations $\bar{x}$ capable of yielding these particular outcomes.
        This can be expressed as
        \begin{align} \label{eq: fidelity specific}
            \mathcal{F}_{A_1,B_1,A_2,B_2} = 
            \sum_{\bar{x}\in \mathcal{S}} p\qty(\bar{x}) \big/ 
            \sum_{\bar{x}\in \mathcal{E}_2} p\qty(\bar{x}) \;.
        \end{align}
        

        The fidelity of the distillation protocol $\mathcal{F}_{\text{avg}}$ is obtained by averaging over all possible measurement outcomes, i.e.,     
        \begin{align}
            \mathcal{F}_{\text{avg}} =
            \sum_{\mathclap{A_1,B_1,A_2,B_2}}
            \mathcal{F}_{A_1,B_1,A_2,B_2} 
            \cdot
            p_{A_1,B_1,A_2,B_2}
            \;.
        \end{align}
        Here, $p_{A_1,B_1,A_2,B_2}$ is the probability of a specific measurement outcome. Since each error $\bar{x}$ has $\frac{m_i}{m_c}$ different outcomes for the angular measurements and $\frac{m_c}{m_f}$ for the photon number measurements, we obtain the following identity:
        \begin{align}
            p_{A_1,B_1,A_2,B_2} = 
            \frac{m_f}{m_i}
            \sum_{\bar{x}\in \mathcal{E}_2} p\qty(\bar{x}) \;.
        \end{align}

    If the protocol abort step is included, i.e., $f_{\text{cut}}>0$, we restrict our attention to only those measurement outcomes for which the fidelity exceeds $f_{\text{cut}}$. In this case, the average fidelity is computed using only those outcomes ${A_1,B_1,A_2,B_2}$ that satisfy $\mathcal{F}_{A_1,B_1,A_2,B_2}>f_{\text{cut}}$:
    \begin{align} \label{eq:fidelity_average_with_cut}
         \mathcal{F}_{\text{avg}} =
        \sum_{\mathclap{\substack{A_1,B_1,A_2,B_2 \\ \text{s.t.}~ \mathcal{F}\qty(A_1,B_1,A_2,B_2) > f_\text{cut}}}}
        \mathcal{F}_{A_1,B_1,A_2,B_2} 
        \cdot
        p_{A_1,B_1,A_2,B_2}
        \;.
    \end{align}

    This also introduces a failure probability, $p_{\text{abort}}$, expressed as
    \begin{align}
        p_{\text{abort}} =
        \sum_{\mathclap{\substack{A_1,B_1,A_2,B_2 \\ \text{s.t.}~ \mathcal{F}\qty(A_1,B_1,A_2,B_2) \le f_\text{cut}}}}
        p_{A_1,B_1,A_2,B_2} \;.
    \end{align}

\twocolumngrid

    \section{No-communication protocol}
    \label{sec: no-communication}

        \begin{figure*}[t] 
        \begin{minipage}{\dimexpr0.5\linewidth\relax} 
            \includegraphics[width=\linewidth]{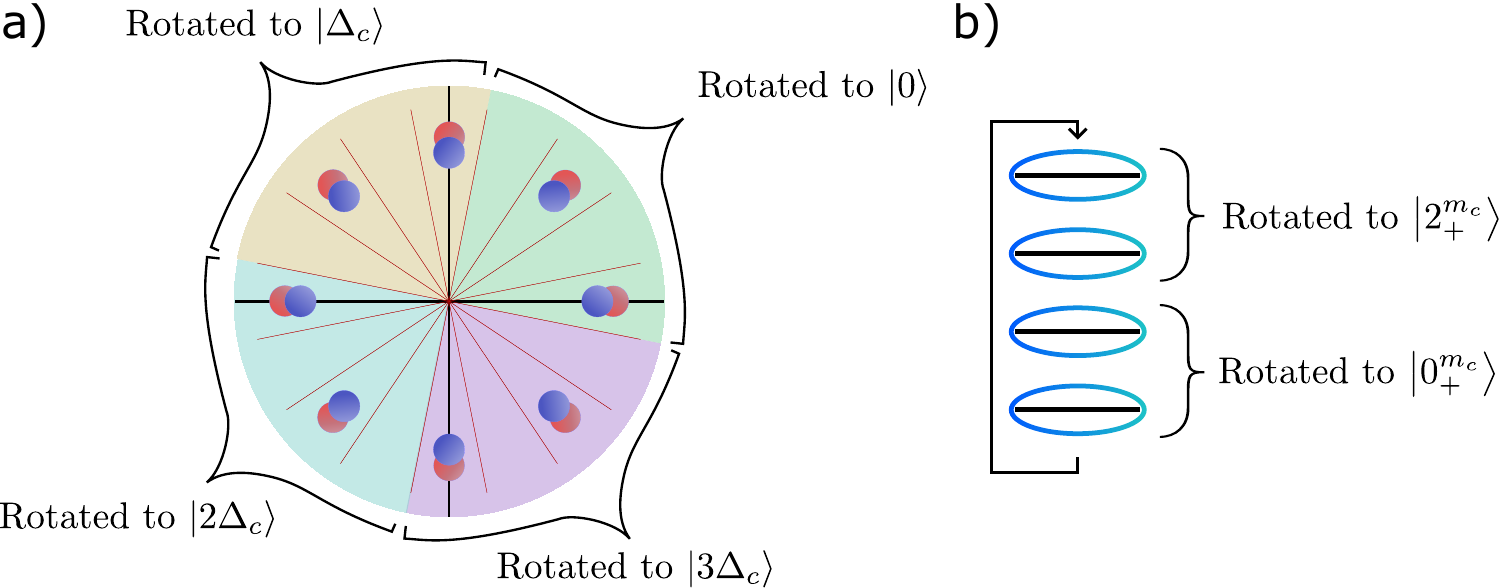}
        \end{minipage}
        \centering
        \caption{Illustration of a no-communication strategy, represented in phase space \textbf{(a)} and in modular Fock space \textbf{(b)}. In this case, $d=16$, $m_i=8$, and $m_c=4$. The blue and red circles represent the states of Alice and Bob prior to the modular phase measurements. After the angular measurement of $O_1 \in \qty{0,...,\Delta_c-1}$, one optimal strategy is to rotate the resulting state clockwise by $-O_1$. In other words, the rotation function is $f(O_1)=0$, meaning we always rotate to the closest alignment clockwise. The dual basis rotation function is $g(O_2)=0$, meaning we always rotate to the closest alignmed state with fewer photons. 
        The no-communication strategy can be thought of as a coarse-graining of the cavity phase space. For example, in the computational basis, measurement outcomes ranging from $\lfloor 1-\Delta_i/2 \rfloor$ to $\lfloor \Delta_c-\frac{1}{2} \Delta_i\rfloor$ are all merged to the same state.}
        \label{fig: no_communication rep}
    \end{figure*}

    In Figure \ref{fig: no-communication} of the main text, we compare the performance of the distillation protocol with a strategy in which there is no communication between the parties. In this section, we describe the no-communication protocol in more detail. After completing a modular measurement in either the computational or dual basis, each party independently applies a phase-space rotation to their respective cavity based solely on their own measurement outcomes ($A_1,A_2$ and $B_1,B_2$ for Alice and Bob, respectively). 

    There are $\Delta_c \qty(\Delta_f)$ possible measurement outcomes and $m_c \qty(m_f)$  different rotations for aligning the state with $\ket{0} \qty(\ket{0_+})$. A map between the outcomes and rotations can be described by a function for each of the two protocol stages, namely $f_O: \mathbb{Z}_{\Delta_c} \rightarrow \mathbb{Z}_{m_c}$ and $g_O: \mathbb{Z}_{\Delta_f} \rightarrow \mathbb{Z}_{m_f}$ with $O\in \qty{A,B}$.


    These functions describe the rotation each party performs as follows:
    \begin{align*} \label{}
        X^{-O_1 + \Delta_c f_O(O_1)}\;,
        \\
        Z^{-O_2 + \Delta_f g_O(O_2)} \;.
    \end{align*}
    These rotations ensure that the state is realigned with $\ket{0}$ and $\ket{0_+^{m_c}}$, respectively. This realignment can be written by applying the rotations to the measured state from Eq. \eqref{eq: collapsed state appendix} after the rotations:
    \begin{multline} \label{eq: rotated no comm state appendix}
            X_A^{-A_1 + \Delta_c f_A(A_1)}
            X_B^{-B_1 + \Delta_c f_B(B_1)}\ket{\tilde{\varphi}_{\bar{x}}} =
            \\
            \frac{1}{\sqrt{m_c}} \sum_{k'=0}^{m_c-1} c_{k'} \ket{\Delta_c (k'+f_A(A_1)}_A \ket{\Delta_c (k' + u + f_B(B_1))}_B ,
    \end{multline} 
    and similarly using Eq. \eqref{eq:state-after-second-collapse-appndx} for the dual basis.
    
    The maps $f_X$, $g_X$ are chosen to maximize the probability for a successful alignment between the cavities, i.e. to maximize the fidelity to the state $\frac{1}{\sqrt{m_c}}\sum_{k'} c_{k'} \ket{\Delta_c k'}_A \ket{\Delta_c k'}_B$.

    Due to the symmetry between the two parties, we assume $f_A=f_B$ and $g_A=g_B$.

    We focus first on the dephasing correction stage.
    Since we assume that large misalignments in the computational basis (i.e., $u>1$) are highly unlikely, the optimal strategy for $f(O)$ is binary: $f(O_1)=0$  corresponds to a clockwise rotation to the nearest aligned configuration, while $f(O_1)=1$ indicates a counter-clockwise rotation to the closest aligned state. 

    Before addressing the general scenario, consider a case where the separation between supporting sections of the initial state is two, i.e., $\Delta_i = d/m_i = 2$ (see Fig. \ref{fig: no_communication rep}).
    In this case, an even measurement outcome $O_1$ is most likely associated with no dephasing error.
    Odd results of $O_1$ are most likely the result of a dephasing rotation by one section. However, the outcome does not convey information on the direction of the dephasing error. 
    Therefore, the optimal strategy is to choose a constant function for $f$, whose value represents a predetermined decision on whether to align the state with $\ket{0}$ or with $\ket{\Delta_c}$.
    
    In the general scenario with $\Delta_i>2$, the measurement outcome can convey information on the direction of the errors. That is, the measurement has a sufficiently high resolution to indicate whether the dephasing error was likely a clockwise or a counterclockwise rotation.


    In this case, it can be shown that the optimal function $f$ changes its value in the middle between two supporting states, i.e. $\lceil(n+\frac{1}{2}) \Delta_i \rceil$ for any $n \in \mathbb{Z}_{m_i/m_c}$. 
    This function can be written as
    \begin{align}
        f_n(O_1) = 
        \begin{cases}
            0 & O_1 \le \lceil\qty(n+\frac{1}{2}) \Delta_i \rceil
            \\
            1 & O_1 > \lceil \qty(n+\frac{1}{2}) \Delta_i \rceil
        \end{cases}
        \;.
    \end{align}
    Note that the specific choice of $n$ is arbitrary due to the translation symmetry of the initial state.

     Following Appendix \ref{sec: analyzing the protocol}, the criterion for a successful realignment is
    \begin{align}
        s_A - s_B = A_1 - B_1 + u_{\text{NC}}\Delta_c \pmod{d} \;,
    \end{align}
    with $u_{NC} = \qty(f(A_1)-f(B_1))$.

    For the loss correction stage, the result $O_2$ does not convey any information on the underlying noise, since all dual-basis states participate equally in the intermediate cavity state. 
    Thus, there is no preferred direction for rotation, and we set $g(O_2)=0$.


    \section{Abort performance for fixed intermediate dimensions} \label{sec: all m_c}

\begin{figure*}[t]
        \centering
        \includegraphics[width=0.8\linewidth]{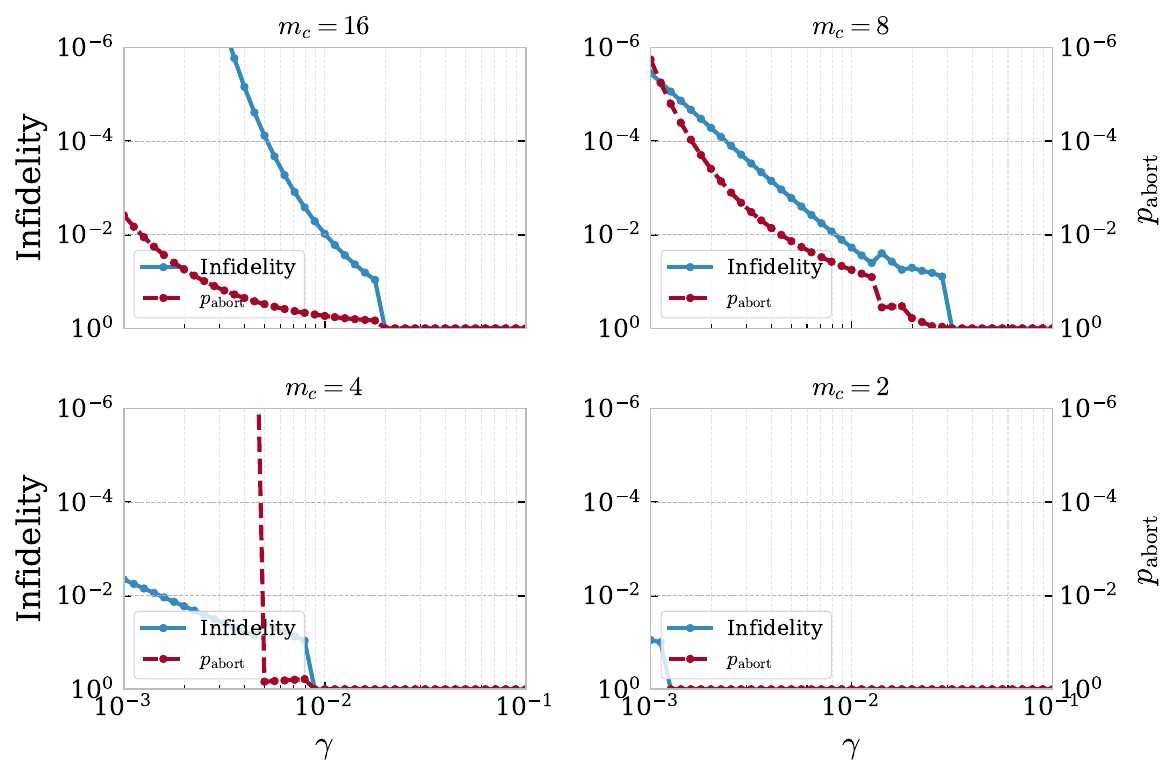}
        \caption{Average protocol infidelity and abort probability for the aborting protocol with fixed values of $m_c$. The parameters are $d=32$, $m_i=16$, $m_f=2$, $\bar{n}=49$, $\gamma_l=\gamma_\phi=\gamma$, and $f_\text{cut}=0.9$, matching Fig.~\ref{fig: aborting} of the main text.}
        \label{fig: all-m_c-aborting}
    \end{figure*}

    In Fig.~\ref{fig: all-m_c-aborting}, we show the infidelity and abort probability for the same parameters used in Fig.~\ref{fig: aborting}, but with fixed values of the intermediate dimension $m_c$.
    These curves separate the behavior of each fixed-$m_c$ protocol from the additional non-monotonic features that can appear when $m_c$ is optimized independently at each physical error rate.

    \section{Distilling higher-entanglement states ($m_f > 2$)} \label{sec: m_f neq two}

    For certain cases, distilled states with higher entanglement entropy may be desirable. 
    In our protocol, this corresponds to setting $m_f > 2$.
    In this section, we evaluate the performance of the protocol for \( m_f = 4 \), which corresponds to an entanglement entropy of \( E_f = 2 \) for the target state.

    Increasing $m_f$ while keeping $d$ and $m_i$ constant reduces the separation between states in either the dephasing correction stage or the loss correction stage.
    The balance between the performance of these two stages is determined by $m_c$, which is now also confined to a smaller set of values, $m_f \le m_c \le m_i$.

\begin{figure}[t] 
        \begin{minipage}{\dimexpr\linewidth\relax} 
            \includegraphics[width=\linewidth]{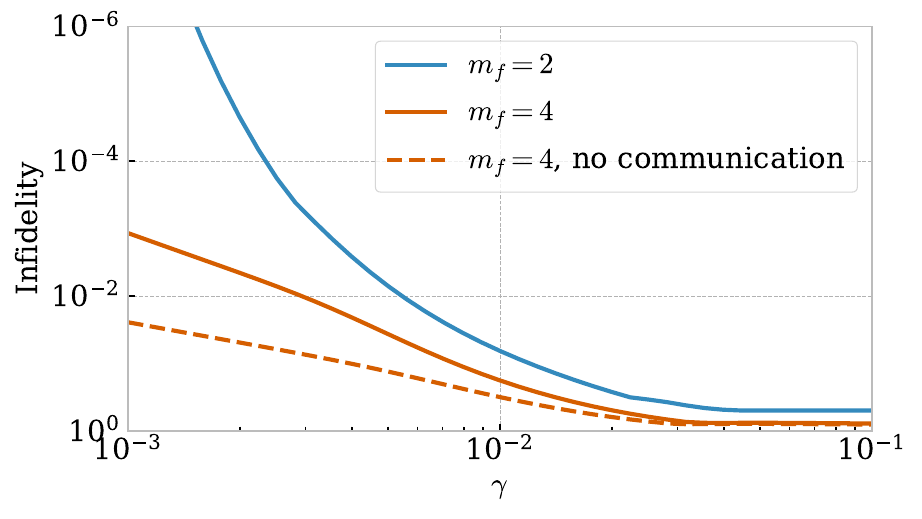}
        \end{minipage}
        \centering
        \caption{Average protocol infidelity as a function of noise coefficients ($\gamma_l = \gamma_\phi=\gamma$) for $m_f=2$ and $m_f=4$. $d=m_i=16$, and $m_c$ is independently optimized for each noise coefficient value.}
        \label{fig: m_f neq 2}
    \end{figure}

    Figure \ref{fig: m_f neq 2} compares the performance of the protocol for $m_f=2$ and $m_f=4$, with the optimal $m_c$ selected for each case.
    Although a higher $m_f$ compromises the protocol performance, a high distillation fidelity is still achieved compared to the no-communication protocol. 

    \section{Distillation of initial states with low entanglement entropy} \label{sec: entanglement resource}

    When analyzing the distillation protocol, it is not evident that choosing an initial state with higher entanglement entropy than that of the desired target state results in a higher fidelity of the distilled state.
    
    To illustrate this, consider the 4-level qudit case with dephasing noise $\mathcal{N}_D$ as described in Section \ref{sec:four-level-example}. Alice can directly prepare the target state $\ket{\psi_i'} = \ket{\psi_t} = \frac{1}{\sqrt{2}} \qty( \ket{00} + \ket{22})$ instead of the proposed initial state $\ket{\psi_i} = \frac{1}{\sqrt{4}}\qty(\ket{00}+\ket{11}+\ket{22}+\ket{33})$.
    The initial state $\ket{\psi_i'}$ has an entanglement entropy of $E(\ket{\psi_i'})=1$, whereas the proposed state has $E(\ket{\psi_i})=2$.
    The state goes through the noise channel, and its fidelity degrades.
    When applying the protocol to the noisy state $\mathcal{N}_D(\todm{\psi'})$, the same fidelity is obtained as for the case of the proposed state $\ket{\psi_i}$. 
    
    However, this property is not unique to our distillation protocol. In traditional qubit-based distillation protocols, one can also identify initial states with low entanglement entropy yielding identical distillation performance to the typical highly-entangled initial state consisting of multiple Bell pairs.
    For example, we consider the recurrence distillation protocol \cite{Bennett_first} applied to two noisy Bell pairs, i.e., two e-bits, described by the state $W_F \otimes W_F$. Here, $W_F$ is the Werner state
    \begin{multline}
        W_F = \Delta_F \qty(\todm{\Psi^-})
        = F \todm{\Psi^-} 
        \\ + \frac{1-F}{3} (\todm{\Psi^+} + \todm{\Phi^-} + \todm{\Phi^+}) \;,
    \end{multline}
    where $\Delta_F$ is the depolarizing channel with parameter $F$ acting on one qubit of the Bell pair,  and $\qty{\Psi^\pm,\Phi^\pm}$ are the Bell basis states.
    Alternatively, the protocol could be applied to the  1-ebit state $(\Delta_F \otimes \Delta_F) \qty(\ket{00}_A \ket{11}_B+\ket{11}_A \ket{00}_B)$  and obtain the same target state with identical fidelity.
    This fact demonstrates that a higher initial entanglement entropy does not necessarily translate into a distilled state with higher fidelity. However, the initial state with low entanglement entropy is harder to produce, and so starting with multiple Bell pairs is preferable.

    The same principle applies to our bosonic entanglement distillation protocol. As described in Section \ref{section: Creating the initial state}, the state $\ket{\psi_i}$ is easier to produce than $\ket{\psi_i'}$, as it requires shorter gate times.
    Moreover, when both loss and dephasing errors are present, preparing the relevant state with low entanglement entropy requires a complex initialization procedure \cite{Grimsmo_2020}, making its preparation even more challenging.


\onecolumngrid
    
    \section{Bosonic operations} \label{sec: bosonic operations}

    In this section, we elaborate on the bosonic continuous-variable operations described in section \ref{sec: bosonic-main}.

    \subsection{Phase measurements}

    Canonical phase measurements are described by the set of POVM's $\qty{M\qty(\theta)}$ where $\theta$ is the measured phase \cite{Holevo_2011,Grimsmo_2020}. The elements of this set can be represented as:
    \begin{align} \label{eq: canonical phase meas}
        M\qty(\theta) = \frac{1}{2\pi} \sum_{m,n=0}^\infty e^{i\qty(m-n)\theta} \ket{m}\bra{n} \;.
    \end{align}

    Since we are interested in measuring whether the phase is in a particular section defined by the interval $\theta \in \qty(\frac{\qty(2k-1)\pi}{d},\frac{\qty(2k+1)\pi}{d})$, we integrate over the phase and obtain the section measurement POVM element
    \begin{align} \label{eq:povm-cannonical}
        M_k &= \int_{\frac{\qty(2k-1)\pi}{d}}^{\frac{\qty(2k+1)\pi}{d}} M\qty(\theta) d\theta
        = \frac{1}{2\pi}\sum_{m,n=0}^{\infty}
        \Gamma_{n,m}^k
        \ket{m}\bra{n} \;,
    \end{align}
    with
    \begin{align}
        \Gamma_{n,m}^k = 
        \begin{cases}
            \frac{1}{i\qty(m-n)}
            \qty(e^{i\qty(m-n)\qty(2k+1)\pi/d}-e^{i\qty(m-n)\qty(2k-1)\pi/d}) & m \neq n
            \\
            \frac{2 \pi}{d} & m=n
        \end{cases}
        \;.
    \end{align}



    \subsection{Primitive states}

    The two primitive states we use in this work are the coherent and the squeezed-coherent states.
    
    The coherent state $\ket{\alpha}$ is defined by acting with the displacement operator on the cavity vacuum state $\ket{0}$, i.e., 
    \begin{align}
        \ket{\alpha} = \mathcal{D}\qty(\alpha)\ket{0} \equiv e^{\alpha \hat{a}^\dagger - \alpha^* \hat{a}}\ket{0}
        \;.
    \end{align}

    To reduce the overlap between primitive states in adjacent sections, squeezed coherent states $\ket{\alpha,\zeta}= \mathcal{D}\qty(\alpha) \mathcal{S} \qty(\zeta) \ket{0} $ can be used instead, where the squeezing operator $\mathcal{S} \qty(\zeta)$ with squeezing parameter $\zeta$ is defined as
    \begin{align}
        \mathcal{S} \qty(\zeta) \equiv e^{\frac{1}{2} \qty(\zeta \hat{a}^{\dagger 2} - \zeta^* \hat{a}^{2})}
        \;.
    \end{align}

    \subsection{Initial state preparation}

    In this section, we describe the preparation of the initial entangled state for the distillation protocol. The basic building block is the controlled rotation gate $\CROT$ defined as 
    


    \[
        \CROT \coloneqq e^{i\frac{2\pi}{m_i} \hat{n}_A \hat{n}_B} \;.
    \]

    The action of the $\CROT$ gate can be understood by considering its effect on a pair of cavities: one initialized in an arbitrary state $\ket{\psi}_A$ and the other in a Fock state $\ket{j}_B$.
    Specifically, the $\CROT_{AB}$ gate performs a rotation on the first cavity by an angle $2\pi j / m_i$, denoted by $R_{2\pi j / m_i}^A$, i.e.,
    \begin{align}
        \CROT_{AB} \ket{\psi}_A\ket{j}_B 
        =
        R_{2\pi j / m_i}^A \ket{\psi}_A\ket{j}_B \;.
    \end{align}
    
    The preparation procedure consists of a channel $\mathcal{C}_x$ acting on the two cavities $A,B$ and an ancilla cavity $C$, each prepared in the primitive state $\ket{0^{\text{CV}}}=\ket{\Theta}$. This channel, illustrated in Fig. \ref{fig: state preparation}, consists of a $\CROT_{AC}$ gate between the first cavity and the ancilla and another $\CROT_{BC}$ gate between the second cavity and the ancilla, followed by a $Z$-basis measurement on the ancilla cavity.
    
    The application of the two entangling gates results in
    \begin{align}
        \CROT_{AC} \CROT_{BC} \qty(\ket{0^{\text{CV}}}_A \ket{0^{\text{CV}}}_B \ket{0^{\text{CV}}}_C)
        & = 
        \CROT_{AC} \CROT_{BC} \qty(\ket{0^{\text{CV}}}_A \ket{0^{\text{CV}}}_B \sum_{k=0}^{m_i-1} \frac{1}{\sqrt{m_i}} \ket{k^{\text{CV}}_+}_C) 
        \\ \label{eq: after two CROTs}
        & =
        \frac{1}{\sqrt{m_i}}
        \sum_{k=0}^{m_i-1} 
        \ket{k^{\text{CV}}}_A \ket{k^{\text{CV}}}_B \ket{k^{\text{CV}}_+}_C
        \\
        & =
        \frac{1}{\sqrt{m_i}}
        \sum_{k=0}^{m_i-1} 
        \ket{k^{\text{CV}}}_A \ket{k^{\text{CV}}}_B \sum_j e^{- 2 i j k \pi / m_i} \ket{j^{\text{CV}}}_C \;.
    \end{align}
    Here, we defined the dual basis states as
    $\ket{k_{+}^{\text{CV}}}  = \frac{1}{\sqrt{\mathcal{N}_k}} \sum_{j=0}^{d-1}e^{-2ijk\pi/d}\ket{j^{\text{CV}}}$, where the normalization coefficient ${\sqrt{\mathcal{N}_k}}=d$ is constant under the approximation that rotated primitive states $\ket{k^{\text{CV}}} \equiv R_{2k\pi/d} \ket{\Theta}$ are mutually orthogonal, 
    i.e., 
    $\bra{0^\text{CV}} \ket{k^{\text{CV}}} = 0$ for every $k\in \mathbb{Z}_d$.
    Each state in this basis is supported only by photon numbers $\ket{n}$ satisfying $n = k \pmod{d}$, leading to the second equality in \eqref{eq: after two CROTs}. We also assumed $m_i=d$.

    Finally, when measuring the ancilla cavity in the logical $Z$-basis, and obtaining the result $x$, we obtain the desired initial state
    \begin{align}
        \mathcal{C}_x \qty(\ket{0^{\text{CV}}}_A\ket{0^{\text{CV}}}_B\ket{0^{\text{CV}}}_E)
        =
        \frac{1}{\sqrt{m_i}} \sum_{k=0}^{m_i-1} e^{-2ikx\pi/m} \ket*{ k^{\text{CV}}}_A\ket*{ k^{\text{CV}}}_B  \;
    \end{align}
up to a phase that can be tracked in software.

    \section{Decoding the continuous-variable distilled state} \label{sec: decoding map}

    To evaluate the fidelity of our final state in the continuous-variable description, we apply a decoding map $\mathcal{D}\otimes \mathcal{D}$ to a two-qudit Hilbert space of dimension $d^2$.

    Here, we opt to implement the decoding process by utilizing the angular-section definition of the $d$-dimensional qudit.
    We do so by using the Pegg-Barnett states \cite{Pegg-Barnett}
        \begin{align}
        \ket{\phi, s} \equiv \frac{1}{\sqrt{s}}
        \sum_{n=0}^{s-1} e^{in\phi} \ket{n}
        \;.
    \end{align}
    which provide an orthonormal basis $\qty{\ket{\phi=2\pi m/s,s}}_{m=0,...,s-1}$ for the truncated $s$-dimensional Fock space. When $s\rightarrow \infty$, the basis states become confined to a particular phase $\phi$ in the bosonic phase space. Since we want our map to be dependent on the angular section, yet indifferent to the angular degree of freedom within the section, we use this basis to trace out the angular degree of freedom inside each section.

    For our purposes, we adopt an alternative representation of the Pegg-Barnett states $\ket{i, \phi}$, associating them with a particular angular section $i$ and angle $\phi$ relative to the section's center. This is expressed by
    \begin{align}
        \ket{i, \phi=2\pi m/s} \equiv \ket{\phi=2\pi(i/d + m/s),s}
    \end{align}
    with $i\in \mathbb{Z}_d$ and $m\in\qty{-s/2d,..., s/2d-1}$. 

    To extract a logical qudit from a continuous variable mixed state $\rho$ (which we assume can be truncated due to its finite energy), the decoding map $\mathcal
    {D}$ traces out the degrees of freedom inside the angular section $\phi$. The matrix elements of the resulting $d$-dimensional qudit density matrix $\sigma$ are given by
    \begin{align}
        \sigma_{ij} = \mathcal{D}(\rho)_{ij} =
        \sum_{m=-s/2d}^{s/2d-1} \bra{i, \phi=2\pi m/s} \rho \ket{j, \phi=2\pi m/s} \;.
    \end{align}
    with $i,j \in \mathbb{Z}_d$.
    
    For a system of two cavities, $A$ and $B$, with density matrix $\rho_{AB}$, we extend the trace-out procedure in a similar way. The elements of the reduced density matrix $\sigma$ in the $d^2$-dimensional two-qudit space are defined by
    \begin{align}
        \sigma_{i_A,i_B,j_A,j_B} = \sum_{m=-s/2d}^{s/2d-1} \sum_{l=-s/2d}^{s/2d-1} 
        \bra{i_A, \phi=2\pi m/s}_A \bra{i_B, \phi=2\pi l/s}_B \rho_{AB} \ket{j_A, \phi=2\pi m/s}_A \ket{j_B, \phi=2\pi l/s}_B \;,
    \end{align}
    with $i_A, i_B,j_A,j_B \in \mathbb{Z}_d$.
    
    Using this reduced density matrix, we evaluate the logical fidelity of the distilled bosonic state with respect to a desired discrete target state $\ket{\psi_t}$ (see Fig. \ref{fig: squeezing effect}). The fidelity conditioned on a specific set of measurement outcomes $(A_1,B_1,A_2,B_2)$ is given by
    \begin{align}
        \mathcal{F} \qty(A_1,B_1,A_2,B_2) = F \qty(\sigma\qty(A_1,B_1,A_2,B_2), \ket{\psi_t} \bra{\psi_t}) \;.
    \end{align}

    \section{Numerical methods}
    \label{sec:numerical-methods}

    The numerical simulations of the continuous-variable protocol are performed using the QuTiP library \cite{qutip5}. The simulations are carried out in a truncated Fock space, with the truncation level chosen to be $N=220$ for the results presented in the main text. This truncation level was chosen based on convergence tests, ensuring that the results are stable and accurate for the range of parameters considered. 

    The simulations are full density-matrix simulations, which allows for the accurate modeling of both loss and dephasing noise channels.

    The measurements are implemented by constructing the corresponding measurement operators as described by Eq.~\eqref{eq:povm-cannonical} for the canonical measurements, and by the operators described in Ref.~\cite{Hillmann_2022arne_noisy_measurements} for the heterodyne and adaptive homodyne measurements.

    For the discrete-variable protocol, the fidelity is calculated directly using Eq.~\eqref{eq:fidelity_average_with_cut}, by summing over all possible measurement outcomes.

    The full codebase is available in \cite{jacoby2026zenodo}.

    \section{Numerical Extensions}

    We present additional numerical results that extend the analysis of the main text.

    \subsection{Break-even behavior for unequal loss and dephasing rates} \label{sec: unequal-loss-dephasing}

    The break-even comparison in Fig.~\ref{fig: squeezing effect} assumes equal loss and dephasing rates, $\gamma_l=\gamma_\phi$. Since the relative magnitudes of these rates can vary across superconducting-cavity hardware and operating conditions, we also evaluate the continuous-variable protocol for fixed ratios $\gamma_l/\gamma_\phi=0.1$ and $\gamma_l/\gamma_\phi=10$.

    \begin{figure}[t]
        \centering
        \includegraphics[width=0.7\linewidth]{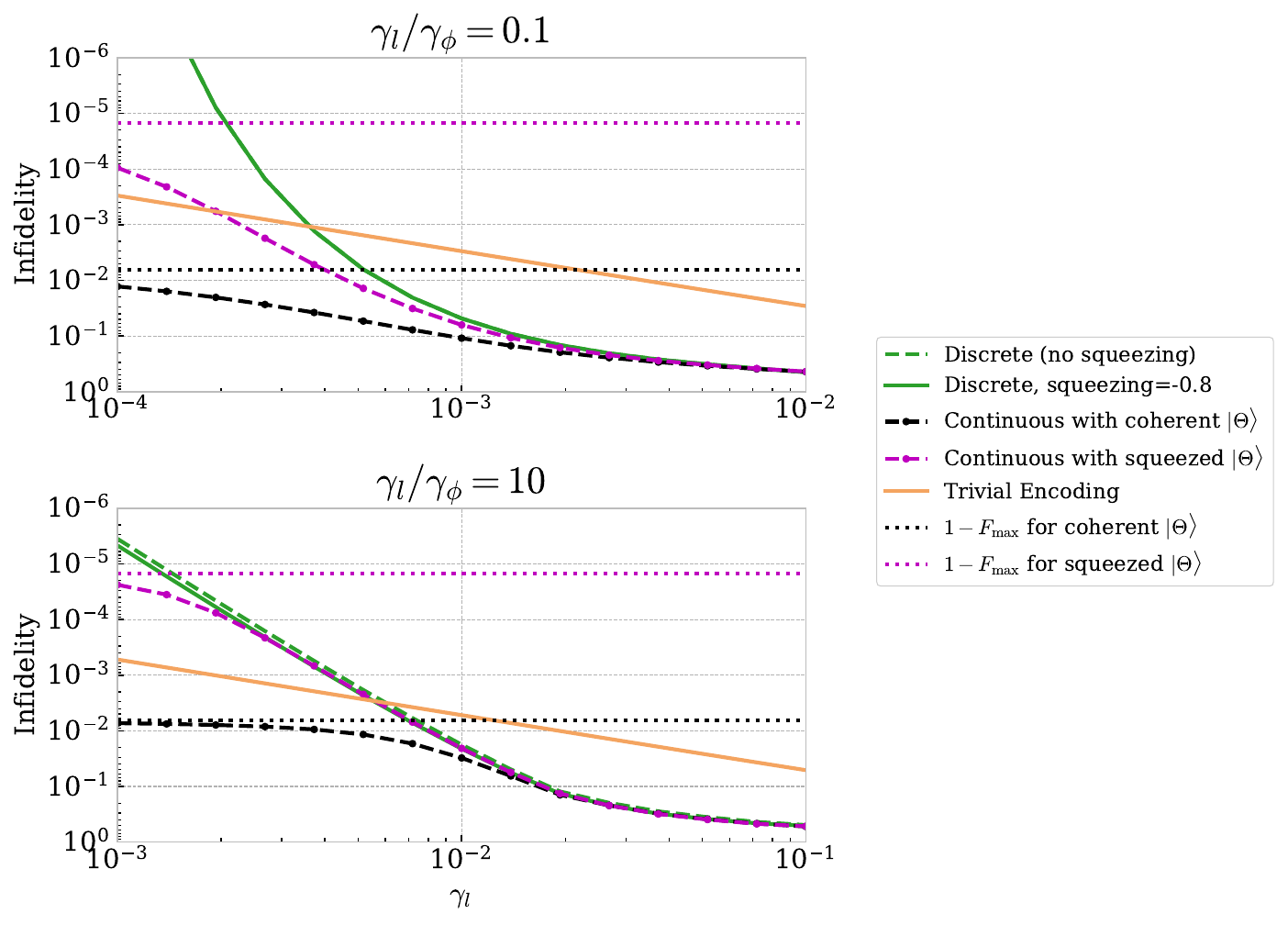}
        \caption{Simulated infidelity of the final distilled state for unequal loss and dephasing rates. We compare the discrete model with the continuous-variable model for coherent and squeezed primitive states, as in Fig.~\ref{fig: squeezing effect}, while fixing $\gamma_l/\gamma_\phi=0.1$ (top) and $\gamma_l/\gamma_\phi=10$ (bottom). The horizontal axis is the loss probability $\gamma_l$, with $\gamma_\phi$ set by the fixed ratio. The dotted horizontal lines indicate the quantum-noise floors for the corresponding continuous-variable primitive states.}
        \label{fig: unequal-loss-dephasing}
    \end{figure}

    \subsection{Quantum noise floor for different $d$ and $m_c$ values} \label{sec:quantum-noise-floor}

\begin{figure}
        \centering
        \includegraphics[width=0.8\linewidth]{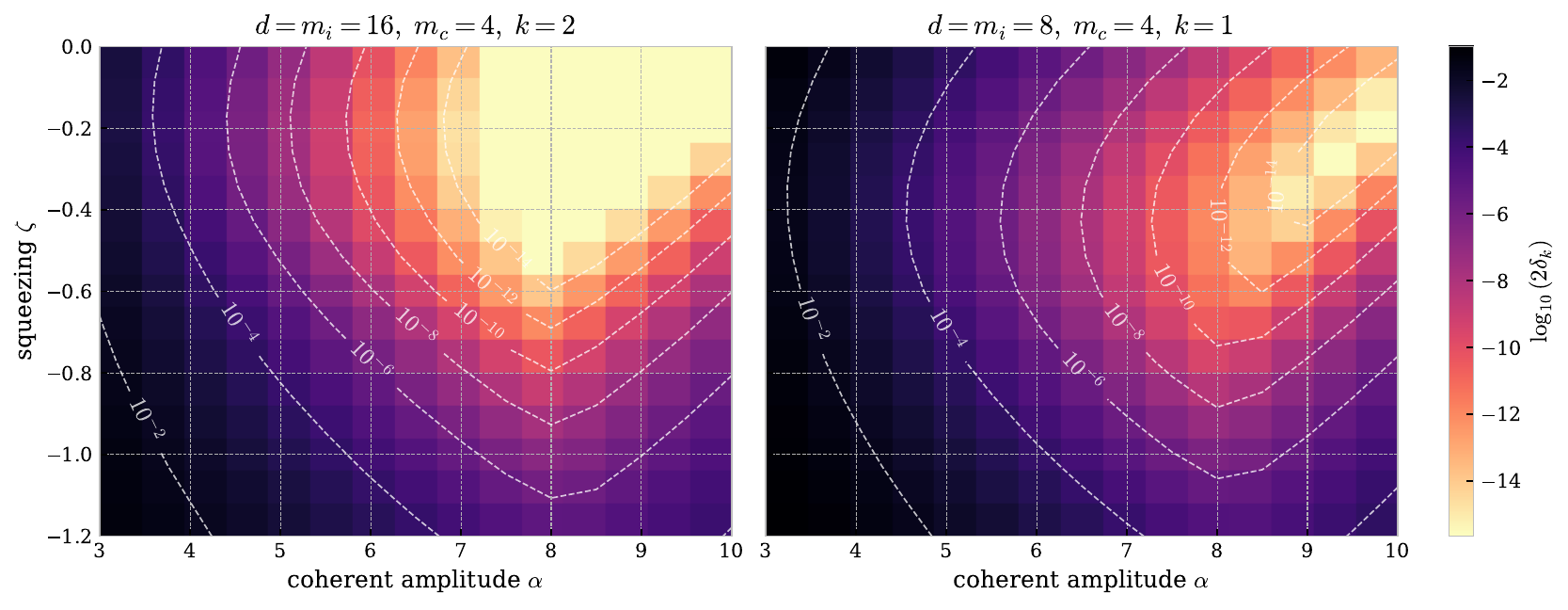}
        \caption{Quantum-noise floor for different values of the initial dimension $d$ and the intermediate dimension $m_c$.}
        \label{fig:quantum-noise-floor-app}
    \end{figure}

    In Fig.~\ref{fig:quantum-noise-floor-app}, we show the quantum-noise floor for different values of the initial dimension $d$ and the intermediate dimension $m_c$.

    \newpage
    \twocolumn
    \bibliographystyle{quantum}
    \bibliography{bib}

\end{document}